\documentclass[prb,preprintnumbers,twocolumn,superscriptaddress,showpacs,nobalancelastpage]{revtex4}

\usepackage{amsmath,amssymb,amsfonts}
\usepackage{bm}
\usepackage{graphicx,color}

\newcommand{\ket}[1]{|{#1}\rangle}
\newcommand{\bra}[1]{\langle{#1}|}
\newcommand{\braket}[2]{\langle{#1}|{#2}\rangle}
\newcommand{\floor}[1]{\lfloor{#1}\rfloor}

\newcommand{\calA}{{\cal A}}
\newcommand{\calB}{{\cal B}}

\newcommand{\calF}{{\cal F}}
\newcommand{\calG}{{\cal G}}
\newcommand{\tone}{t_1}
\newcommand{\tonesq}{t_1^2}
\newcommand{\ttwo}{t_2}

\newcommand{\tonetwo}{t_1\text{-}t_2}
\newcommand{\txpre}{t^{\protect\ast}}
\newcommand{\txpresq}{t^{\protect\ast2}}
\newcommand{\txone}{t_1^{\protect\ast}}
\newcommand{\txonesq}{t_1^{\protect\ast2}}
\newcommand{\txtwo}{t_2^{\protect\ast}}
\newcommand{\txtwosq}{t_2^{\protect\ast2}}
\newcommand{\husimianhang}{the Appendix}

\begin{document}

  \title{Hopping on the Bethe~lattice: Exact~results
    for densities~of~states and dynamical~mean-field~theory}

  \author{Martin Eckstein}
  
  \author{Marcus Kollar}

  \affiliation{Theoretical Physics III, Center for Electronic
    Correlations and Magnetism, Institute for Physics, University of
    Augsburg, D-86135 Augsburg, Germany}

  \author{Krzysztof Byczuk}
  
  \affiliation{Institute of Theoretical Physics, Warsaw University,
    ul.\ Ho\.za 69, PL-00-681 Warszawa, Poland}

  \author{Dieter Vollhardt}
  
  \affiliation{Theoretical Physics III, Center for Electronic
    Correlations and Magnetism, Institute for Physics, University of
    Augsburg, D-86135 Augsburg, Germany}
  
  \date{September 28, 2004}

  \begin{abstract}
\vspace*{3mm}
    We derive an operator identity which relates tight-binding
    Hamiltonians with arbitrary hopping on the Bethe lattice to the
    Hamiltonian with nearest-neighbor hopping. This provides an exact
    expression for the density of states (DOS) of a non-interacting
    quantum-mechanical particle for any hopping.  We present analytic
    results for the DOS corresponding to hopping between nearest and
    next-nearest neighbors, and also for exponentially decreasing
    hopping amplitudes. Conversely it is possible to construct a
    hopping Hamiltonian on the Bethe lattice for any given DOS.  These
    methods are based only on the so-called distance regularity of the
    infinite Bethe lattice, and not on the absence of loops.  Results
    are also obtained for the triangular Husimi cactus, a recursive
    lattice with loops.  Furthermore we derive the exact
    self-consistency equations arising in the context of dynamical
    mean-field theory, which serve as a starting point for studies of
    Hubbard-type models with frustration.
  \end{abstract}

  \pacs{
    71.10.Fd
    ,
    71.27.+a
    ,
    05.50.+q
    ,
    02.10.Ox
\vspace*{4mm}
  }
  \maketitle

  \section{Introduction}\label{sec:intro}
  
\begin{figure}[t]
    \centerline{\includegraphics[clip,height=0.75\hsize,angle=0]{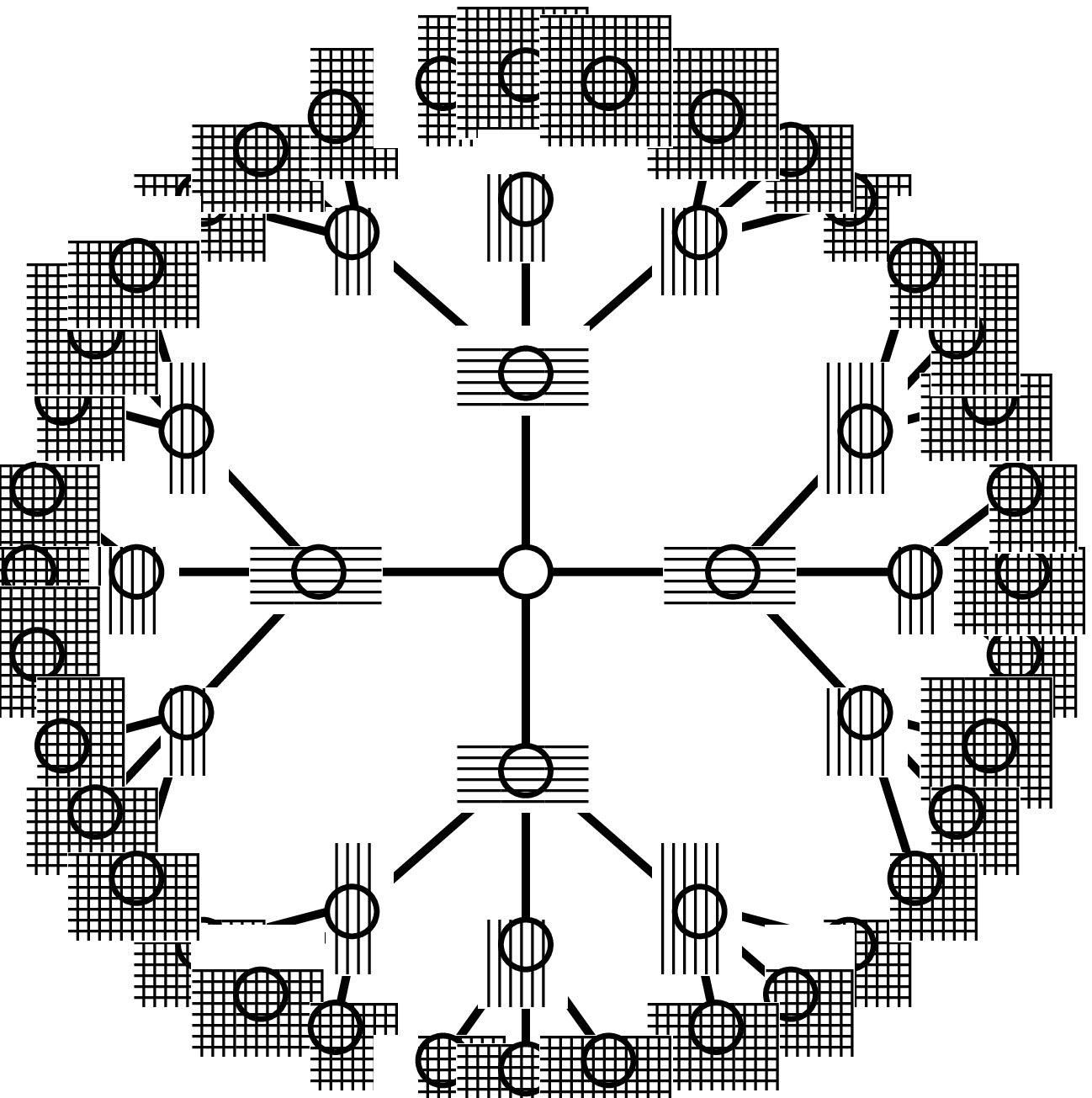}}
    \nopagebreak
    \caption{Part of the Bethe lattice with coordination number $Z=4$. 
      Any two sites are connected by a unique shortest path of bonds.
      Starting from the site marked by the open circle, horizontally
      shaded circles can be reached by one lattice step (NN),
      vertically shaded circles by two lattice steps (NNN), and doubly
      shaded circles by three lattice steps. Note that the lattice is
      infinite and that all sites are equivalent; the shading appears only
      for visualization of hopping processes.}
    \label{fig:bethe}
  \end{figure}

  The Bethe lattice is an infinite graph, where any two points are
  connected by a single path and each vertex has the same number of
  branches $Z$, as shown in Fig.~\ref{fig:bethe} for $Z=4$.  The name
  ``Bethe lattice'' originates from the fact that Bethe's
  approximation for the Ising model is exact on this
  lattice.\cite{bethe,baxter} A finite portion of the Bethe lattice is
  called Cayley tree. The latter has a peculiar thermodynamic limit due to
  its large surface,\cite{baxter,eggarter,muellerhartmann,yndurain}
  whereas the infinite Bethe lattice has no surface, all its
  lattice sites being located inside the infinite tree.
  
  Strictly speaking the Bethe lattice is a pseudolattice because it
  does not possess the usual point and translational symmetries of
  crystal (Bravais) lattices.  Nevertheless, it plays an important
  role in statistical and condensed-matter physics because some
  problems involving disorder and/or interactions can be solved
  exactly when defined on a Bethe lattice, e.g., Ising
  models,\cite{bethe,baxter,hu}
  percolation,\cite{fisher,chalupa,zhang} or Anderson
  localization.\cite{abouchacra,mirlin,efetov,zirnbauer} Such exact
  solutions on the Bethe lattice for $Z$ $<$ $\infty$
  sometimes,\cite{bethe,baxter} but not
  always,\cite{abouchacra,mirlin,efetov,zirnbauer} have mean-field
  character.  Furthermore, it was argued that mean-field theories are
  more reliable if derived on a Bethe lattice.\cite{gujrati} On the
  other hand, the Bethe lattice may actually serve as a model for the
  electronic structure of amorphous solids, as proposed by Weaire and
  Thorpe,\cite{weaire} e.g., for hydrogenated amorphous
  silicon.\cite{verges} Laughlin and Joannopoulos\cite{laughlin} used
  the Bethe lattice to describe lattice vibrations in amorphous
  silica, and recently this approach was also applied to phonon
  transport through silica-coated nanowires.\cite{mingo}
  
  There are two special properties that make the Bethe lattice
  particularly suited for theoretical investigations. One is its
  self-similar structure which may lead to recursive solutions.  The
  other is the absence of closed loops which restricts interference
  effects of quantum-mechanical particles in the case of
  nearest-neighbor (NN) coupling.  The situation is different if also
  longer-range hopping processes or interactions are allowed, e.g.,
  between next-nearest neighbors (NNN). For example, the frustration
  introduced by NNN hopping typically suppresses antiferromagnetism in
  the half-filled Hubbard model at weak coupling.
  
  In this paper we consider tight-binding Hamiltonians describing
  hopping of a single quantum particle, paying special attention the
  case where the hopping has both NN and NNN contributions (with
  respective amplitudes $\tone$, $\ttwo$), and to the limit $Z$ $\to$
  $\infty$.  The derivation of the spectrum for NN hopping has a long
  history involving many different
  methods.\cite{brinkman-rice,chen,economou,thorpe,mahan} However,
  these methods are not immediately useful for longer-range hopping.
  Here we develop a method, based on the algebraic properties of
  so-called distance-regular graphs, which can effectively treat
  arbitrary-range hopping on the Bethe lattice and certain other
  lattices.  In particular, we derive an operator identity
  {[Eq.~(\ref{eq:geometricseries})]} relating tight-binding
  Hamiltonians $H_d$, which describe hopping between sites that are
  $d$ NN steps apart, to $H_1$, the Hamiltonian for NN hopping.  This
  general result has several applications.  For comparison with
  earlier methods we first note that the nonlocal Green function for
  NN hopping can be obtained via a rather short route.  We proceed to
  derive the exact density of states (DOS) for arbitrary hopping and
  discuss in detail the case of $\tonetwo$ hopping and exponentially
  decreasing hopping.  The inverse problem, i.e., the construction of
  a tight-binding Hamiltonian on the Bethe lattice corresponding to a
  \emph{given} DOS, is also solved.
  
  An important limit for any lattice is that of infinite coordination
  number, $Z$ $\to$ $\infty$, since it always leads to a mean-field
  theory of some sort.\cite{baxter} It is well known that for
  fermionic lattice models the hopping matrix elements must then be
  properly rescaled, e.g., $\tone\sim1/\sqrt{Z}$ for NN
  hopping.\cite{vollhardt} In this limit dynamical mean-field theory
  (DMFT)\cite{Georges92,Jarrell92,vollha93,pruschke,georges,PT}
  becomes exact, yielding self-energies that are local in space.  In
  particular the Hubbard model may be mapped onto a single-impurity
  problem with self-consistency condition.\cite{georges} Here the
  Bethe lattice leads to further simplifications, e.g., in the
  solution of the self-consistency equations, partly due to the
  resulting semielliptical density of states for NN hopping,
  $\rho(\epsilon)$ $=$ $2\sqrt{1-(\epsilon/D)^2}/(\pi D)$, where $D$
  is the half-bandwidth.\cite{economou,thorpe} Interestingly the
  algebraic band edges of this model DOS resemble those of
  three-dimensional systems.  DMFT with this DOS has been
  instrumental in clarifying the phase diagram of the Hubbard model,
  in particular concerning the Mott transition from a paramagnetic
  metal to a paramagnetic insulator at
  half-filling.\cite{Georges92,Jarrell92,vollha93,pruschke,%
    georges,PT,bulla99,Rozenberg,Joo,Bulla01,blumer}
  
  It should be noted that the Mott transition in the Hubbard model
  with NN hopping is usually hidden by an antiferromagnetic
  low-temperature phase,\cite{Jarrell92,rozenberg2,georges} whereas
  for transition metal oxides such as V$_2$O$_3$ the Mott transition
  line extends beyond the antiferromagnetic phase
  boundary.\cite{mcwhan} In order to describe this experimental phase
  diagram at least qualitatively in terms of a Hubbard model, the
  strong tendency towards antiferromagnetism must be reduced, e.g., by
  including the ever-present NNN hopping $\ttwo$.  This was
  demonstrated within a DMFT setup in which the DOS remains
  semielliptic, leading to a suppression of antiferromagnetism
  without modification of the paramagnetic
  phase;\cite{rozenberg2,rozenberg,zitzler,georges} the results of
  this setup are valid for \emph{random} hopping on the Bethe
  lattice.\cite{rozenberg2,georges} In the present work we consider
  standard tight-binding hopping \emph{without randomness}; we find in
  particular that for $\tonetwo$ hopping on the Bethe lattice the DOS
  is no longer semielliptic but becomes asymmetric. We also evaluate
  the exact DMFT self-consistency equations for arbitrary hopping on
  the Bethe lattice, including phases with broken sublattice symmetry.
  Our results for $\tonetwo$ hopping [Eq.~(\ref{eq:weiss-ttprime-AB})]
  differ from the corresponding equations employed in
  Refs.~\onlinecite{georges,rozenberg2,rozenberg,zitzler}; therefore
  the latter only apply to \emph{random} hopping.
  
  The paper is structured as follows: In Sec.~\ref{sec:general} we
  consider topological aspects of tight-binding Hamiltonians.  In
  Sec.~\ref{sec:opident} a general operator identity for the Bethe
  lattice is derived, which is used in Sec.~\ref{sec:dos} to obtain
  the DOS for various hopping ranges; \husimianhang
  contains similar results for the triangular Husimi cactus. Hopping
  amplitudes for a given DOS are constructed in
  Sec.~\ref{subsec:arbdos}.  In Sec.~\ref{sec:dmft} the DMFT
  self-consistency equations are derived.  Our results are discussed
  in Sec.~\ref{sec:conclusion}. Throughout the paper we make contact with
  particular results for the DOS, Green functions, and DMFT
  self-consistency equations which were previously obtained by other
  methods;\cite{blumer,radke,tanaskovic} a detailed comparison will be
  discussed in a separate publication.\cite{kollar05a}

  \section{Tight-binding Hamiltonians and topological lattice properties}\label{sec:general}
  
  In this section we discuss relations among tight-binding
  Hamiltonians for lattices that belong to a certain class of graphs.
  In general a graph $G$ is defined as a set of vertices
  $V=\{i,j,k,...\}$ representing nodes or sites of a network, and a
  set of links between these vertices.  In particular, any Bravais
  lattice is a graph with vertices and links corresponding to lattice
  sites and bonds between nearest neighbors, respectively.  For a
  given lattice one distinguishes between the \emph{metric} and
  \emph{topological} distance between two vertices $i$ and $j$.  The
  metric distance is determined by the metric properties of the space
  in which a graph is embedded.  On the other hand, the topological
  distance between sites $i$ and $j$, denoted by $d_{ij}$ hereafter,
  is the smallest number of links joining $i$ and $j$.  In this paper
  we only use the topological properties and distances of the Bethe
  lattice. Note that the Bethe lattice can be embedded in a hyperbolic
  (Lobachevsky) space with metric properties different from Euclidian
  space.\cite{miranda}
  
  We now consider hopping Hamiltonians on an arbitrary, infinite graph; for
  simplicity we assume at most one link between two sites and no loops
  of length one.  In terms of the quantum-mechanical single-particle
  operator $\ket{i}\bra{j}$, which removes a particle from site $j$
  and recreates it at site $i$, the general tight-binding hopping
  Hamiltonian can be written as
  \begin{subequations}%
    \begin{align}%
      H
      &=
      \sum_{i,j\in V}
      t_{ij} \ket{i}\bra{j}
      =
      \sum_{d\geq0}
      t_{d} H_{d}
      \,,\label{eq:H-general}
    \end{align}%
    where
    \begin{align}%
      H_{d}
      &=
      \sum_{\substack{i,j\in V\\d_{ij}=d}}
      \ket{i}\bra{j}
      \label{eq:H-nu}
    \end{align}%
    \label{eq:H}%
  \end{subequations}%
  describes hopping between sites $i$ and $j$ separated by topological
  distance $d$, i.e., $t_{ij}$ $=$ $t_{d_{ij}}$. By definition all
  nonzero matrix elements of $H_1$ are equal to 1, i.e.,
  \begin{align}
    (H_1)_{ij}
    =
    \bra{i}H_1\ket{j}
    =
    \left\{
      \begin{array}{ll}
        1&~\text{if~}d_{ij}=1\,,
        \\
        0&~\text{otherwise}\,,
      \end{array}
    \right.
  \end{align}
  and $H_0$ $=$ $\sum_i\ket{i}\bra{i}$ $=$ $\openone$ is the
  identity.
  
  The topological properties of a graph are completely described by
  the nearest-neighbor hopping Hamiltonian $H_1$.  This concept is
  used in graph theory where $H_1$ is called {\em adjacency
    matrix}.\cite{graphtheory} Furthermore, a simple interpretation
  can be given to the matrix elements of higher powers of the hopping
  Hamiltonian $H_1$, i.e., $(H_1^n)_{ij}$ is {\em the number of paths
    connecting sites $i$ and $j$ in $n$ NN steps}.\cite{graphtheory}
  Explicitly, one has
  \begin{align} 
    {(H_1^n)}_{ij}
    &=
    \sum_{k_1,\ldots k_{n-1} \in V}
    (H_1)_{ik_{1}}
    (H_1)_{k_{1}k_{2}}
    \cdots
    (H_1)_{k_{n-1}j}
    \,,\label{eqn4}
  \end{align}
  where a term in the sum on the right hand side is equal to unity if
  the string of indices ($ik_1k_2\cdots k_{n-1}j$) represents a path
  joining $i$ and $j$, and is zero otherwise.  Hence, ${(H_1^n)}_{ij}$
  is equal to the number of paths connecting $i$ and $j$ with $n$ NN
  steps.
  
  On the Bethe lattice the quantity ${(H_1^n)}_{ij}$ turns out to be a
  function only of $n$ and of the topological distance $d_{ij}$, while
  the specific positions of $i$ and $j$ are unimportant.  Graphs with
  this property are called {\em distance regular} in graph
  theory.\cite{fiol} The triangular Husimi cactus, which is a set of
  triangles connected by vertices (see \husimianhang), is
  also distance-regular, and hence this property does \emph{not}
  depend on the absence of loops.  However, the situation is different
  for periodic (Bravais) lattices.  For example, given a site $i$ on
  the two-dimensional square lattice, there are two different
  positions for a next-nearest-neighbor site ($d_{ij}=2$): one along
  an axis, the other on the diagonal of a plaquette. In the former
  case there is only one path of length two between $i$ and $j$,
  whereas in the latter case there are two paths.  Thus for crystal
  lattices not only topological distances but also specific positions
  play a role in determining $H_1^n$.
  
  For distance-regular graphs $H_d$ can be written in terms of powers
  of $H_1$.\cite{fiol} We show this by first proving the inverse
  relation, i.e.,
  \begin{align}
    (H_1)^n
    &=
    \sum_{d=0}^n
    a_n^{(d)}\;
    H_d
    \,,\label{eq:a-trafo}
  \end{align}%
  where $a_n^{(d)}$ is the number of paths with $n$ NN steps between
  sites separated by a topological distance $d$.  This relation can be easily
  verified by calculating the corresponding matrix elements
  \begin{align}
    (H_1^n)_{ij}=\sum_{d=0}^n a_n^{(d)} (H_d)_{ij}.
  \end{align}
  Using $(H_d)_{ij}=\delta_{d,d_{ij}}$ one finds
  $(H_1^n)_{ij}=a_n^{(d_{ij})}$, which indeed agrees with
  Eq.~(\ref{eqn4}), hence proving (\ref{eq:a-trafo}).  Since
  $a_n^{(n)}$ $\neq$ $0$ for all $n$, and $a_n^{(d)}$ $=$ $0$ if $d$
  $>$ $n$, this (triangular) system of equations can always be
  inverted, yielding
  \begin{align}
    H_d
    &=
    \sum_{n=0}^d
    A_d^{(n)}\;
    (H_1)^n
    \,.\label{eq:A-trafo}
  \end{align}%
  Therefore for distance-regular graphs {\em the hopping Hamiltonians
    $H_{d}$ are given by polynomials in $H_1$ of order $d$}.  For
  example, for a Bethe lattice with coordination number $Z$ the first
  few equations (\ref{eq:a-trafo}) read
  \begin{subequations}%
    \begin{align}%
      (H_1)^2 &= Z\openone + H_2
      \,,\label{eq:twostep}\\
      (H_1)^3 &= (2Z-1)H_1 + H_3
      \,,\\
      (H_1)^4 &= Z(2Z-1)\openone + (3Z-2)H_2 + H_4
      \,.
    \end{align}%
  \end{subequations}%
  By contrast, such relations do not exist for all path lengths $n$
  and topological distances $d$ on periodic lattices, although
  low-order relations may be found for certain graphs.  For example
  for the honeycomb lattice, which is similar to a Bethe lattice with
  $Z$ $=$ $3$ when taking at most two NN steps, $H_1^2$ can be related
  to $H_1$ and $H_2$ as in Eq.~(\ref{eq:twostep}).

  \section{Operator identities for Bethe lattices}\label{sec:opident}
  
  We now determine the coefficients in Eqs.~(\ref{eq:a-trafo}) and
  (\ref{eq:A-trafo}) for the Bethe lattice.
  These equations can be summarized as an operator identity involving
  the hopping Hamiltonians $H_d$. As an application the Green function
  for NN hopping is obtained.

  \subsection{Recursive relations and generating function}\label{subsec:genfunc}
  
  We consider a Bethe lattice with coordination number $Z$ $\geq$ $2$,
  i.e., branching ratio $K$ $=$ $Z-1$ $\geq$ $1$. In this case the
  coefficients $a_n^{(d)}$ in (\ref{eq:a-trafo}) are obtained from a
  simple recursion, starting from $a_0^{(0)}$ $=$ $a_1^{(1)}$ $=$ $1$.
  Namely, each path of length $n$ $\geq$ $1$ joining a site $i$ to a
  different site $j$ (with $d$ $=$ $d_{ij}$ $\geq$ $1$) is composed of
  a path that joins $i$ to some nearest neighbor of $j$ within $n-1$
  steps and one final step to $j$. Since for the last step there is
  only one possibility, $a_n^{(d)}$ is given by the number of possible
  paths with $(n-1)$ steps between $i$ and the nearest neighbors of
  $j$. Of the $Z$ nearest-neighbor sites of $j$, $Z-1$ are separated
  from $i$ by a distance $d+1$ and one by distance $d-1$.  Together
  with a similar argument for $d$ $=$ $0$ we thus have the recurrence
  relations
  \begin{subequations}%
    \begin{align}%
      a_n^{(d)}
      &=
      Ka_{n-1}^{(d+1)}+a_{n-1}^{(d-1)}
      \,,~~
      d\geq1
      \,,
      \\
      a_n^{(0)}
      &=
      Za_{n-1}^{(1)}
      \,,
      \\
      a_n^{(n)}
      &=
      1
      \,,
    \end{align}%
    \label{eq:a-recursion}%
  \end{subequations}%
  where the last equation is due to the treelike structure of the
  Bethe lattice.  Since on the Bethe lattice it is impossible to
  return to the same site within an odd number of steps, we note that
  $a_n^{(d)}$ vanishes if $n+d$ is odd.
  
  These recursion relations for the Bethe lattice can be solved in
  closed form.  One first considers the generating functions $F_d(u)$
  $=$ $\sum_{n=0}^{\infty}a_n^{(d)}u^n$, which appear when summing
  over Eq.~(\ref{eq:a-trafo}), i.e., $[1-uH_1]^{-1}$ $=$
  $\sum_{d=0}^{\infty}H_dF_d(u)$. They obey the recursion relations
  $F_{d}(u)$ $=$ $KuF_{d+1}(u)+uF_{d-1}(u)$ for $d$ $\geq$ $1$ and
  $F_{0}(u)$ $=$ $1+uZF_1(u)$. These equations may be solved by the
  ansatz\cite{riordan} $F_d(u)$ $=$ $f(u)\,[u\,g(u)]^d$, which yields
  $g(u)$ $=$ $2/[1+\sqrt{1-4Ku^2}]$ and $f(u)$ $=$
  $[1-Zu^2g(u)]^{-1}$, where the sign in front of the square root has
  been chosen so that $F_d(u)$ $=$ $u^d(1+O(u))$ is satisfied.  By
  setting $x$ $=$ $ug(u)$ and rearranging the terms we finally obtain the
  remarkable operator identity
  \begin{align}
    \frac{1-x^2}{1-xH_1+Kx^2}
    &=
    \sum_{d=0}^{\infty}H_d\,x^d
    \,.\label{eq:geometricseries}
  \end{align}
  From this formal power series the coefficients in Eqs.~(\ref{eq:a-trafo})
  and (\ref{eq:A-trafo}) can be extracted in closed form.
  After some algebra we find that the nonzero coefficient $a_n^{(d)}$
  and $A_n^{(d)}$ are given by
  \begin{align}
    a_n^{(n-2s)}
    &=
    \sum_{r=0}^s
    \left[
      \binom{n}{r}
      -
      \binom{n}{r-1}
    \right]
    K^r
    \,,\label{eq:a-result}
    \\
    A_{n}^{(n-2s)}
    &=
    (-1)^s
    \sum_{r=0}^1
    \binom{n-s-r}{s-r}\,K^{s-r}
    \,,\label{eq:A-result}
  \end{align}
  for $0$ $\leq$ $2s$ $\leq$ $n$.  By induction with respect to $n$ it
  is also straightforward to verify that Eq.~(\ref{eq:a-result}) is
  the solution of the recursion relations (\ref{eq:a-recursion}).
  
  Note that in the limit of infinite connectivity, $K$ $\to$ $\infty$
  (or $Z$ $\to$ $\infty$), one must scale\cite{vollhardt} $H_d$ by
  $Z_d^{-1/2}$, where $Z_d$ is the number of sites that a given site
  is connected to by $H_d$. To leading order one has $Z_d$ $\sim$
  $Z^d$ $\sim$ $K^d$; thus we introduce scaled operators,
  $\tilde{H}_d$ $:=$ $H_d/K^{d/2}$.  By computing coefficients we then
  obtain from Eq.~(\ref{eq:geometricseries}) the relation
  \begin{subequations}%
    \begin{align}%
      \tilde{H}_d
      =
      U_d(\tilde{H_1}/2)
      &-\frac{1-\delta_{d0}-\delta_{d1}}{K}\,U_{d-2}(\tilde{H}_1/2)
      \,,\label{eq:chebychev1}
      \\
      U_n(\tilde{H_1}/2)
      &=
      \sum_{s=0}^{\floor{n/2}}
      \frac{\tilde{H}_{n-2s}}{K^s}
      \,,\label{eq:chebychev2}
    \end{align}%
    \label{eq:chebychev}%
  \end{subequations}%
  which are valid for any $K$ $\geq$ $1$ and reduce to $\tilde{H}_d$
  $=$ $U_d(\tilde{H}_1/2)$ in the limit $K$ $\to$ $\infty$.  Here
  $U_n(x)$ are the Chebyshev polynomials of the second
  kind,\cite{abramowitz84a} with generating function
  $[1-2xt+t^2]^{-1}$ $=$ $\sum_{n=0}^{\infty}U_n(x)t^n$.  It seems
  that the operator identities (\ref{eq:geometricseries}) and
  (\ref{eq:chebychev}) might also be useful in other contexts
  involving the Bethe lattice. Note that these relations remain valid
  for the one-dimensional chain ($K$ $=$ $1$).  For the triangular
  Husimi cactus a relation similar to (\ref{eq:geometricseries}) is
  derived in \husimianhang.

  \subsection{Green function}\label{subsec:greenfunc}
  
  As a first application of the operator identity
  (\ref{eq:geometricseries}) we present a shortcut to compute the
  Green function for the NN hopping Hamiltonian $\tone H_1$ on the
  Bethe lattice,
  \begin{align}
    G_{ij}(z)
    &=
    \bra{i}
    (z-\tone H_1)^{-1}
    \ket{j}
    \,,\label{eq:resolvent}
  \end{align}
  with $\text{Im}\,z$ $\neq$ $0$. This will fit into
  Eq.~(\ref{eq:geometricseries}) if we set $z$ $=$ $\tone (1+Kx^2)/x$,
  i.e.,
  \begin{align}
    x
    &=
    \frac{2\tone }{z+\sqrt{z^2-4K\tonesq }}
    \,,
  \end{align}
  with $\text{sgn}(\text{Im}\,\sqrt{z^2-4K\tonesq })$ $=$
  $\text{sgn}(\text{Im}\,z)$.  Since for the Bethe lattice there is
  only one non-self-intersecting path connecting sites $i$ and $j$ we
  have $\bra{i}H_d\ket{j}$ $=$ $\delta_{d,d_{ij}}$, whence
  \begin{multline}
    G_{ij}(z)
    =
    \bra{i}
    \left[
    \frac{1}{\tone }
    \frac{x}{1-x^2}\sum_{d=0}^{\infty}H_dx^d
    \right]
    \ket{j}
    =
    \frac{1}{\tone }
    \frac{x^{d_{ij}+1}}{1-x^2}
    \\
    =
    \frac{2K}{(K-1)z+Z\sqrt{z^2-4K\tonesq }}
    \Bigg[
      \frac{2\tone }{z+\sqrt{z^2-4K\tonesq }}
    \Bigg]^{d_{ij}}
    \!\!\,,\label{eq:G1-nonlocal}
  \end{multline}
  in agreement with Refs.~\onlinecite{economou} and
  \onlinecite{thorpe}. This yields the well-known expression for the
  DOS of $\tilde{H}_1$,
  \begin{align}
    \rho_1(\lambda)
    &=
    -\frac{1}{\pi}\text{Im}\,G_{ii}(\lambda+i0)
    =
    \frac{1}{2\pi}
    \frac{\sqrt{4-\lambda^2}}{p-\lambda^2/Z}
    \,,\label{eq:rho1}
  \end{align}
  where we set $|\tone |$ $=$ $1/\sqrt{K}$ and $p$ $=$ $Z/K$ $=$ $1+1/K$.
  In the limit $Z$ $\to$ $\infty$ this leads to the familiar
  semielliptic density of states,
  \begin{align}
    \rho^{\infty}_1(\lambda)
    :=
    \lim\limits_{K\to\infty}
    \rho_1(\lambda)
    =
    \frac{\sqrt{4-\lambda^2}}{2\pi}
    \,,\label{eq:rho1infZ}
  \end{align}
  as mentioned in the introduction, with half-bandwidth $D$ $=$ $2$.
  For brevity, here and below we use the convention that any square
  root that appears in a DOS yields zero if its argument
  is negative.

  \section{Density of states for arbitrary-range hopping}\label{sec:dos}
  
  The DOS encodes information about hopping parameters and underlying
  lattice structure.  For general tight-binding Hamiltonians on the
  Bethe lattice we now derive an expression for the DOS by using the
  operator identities of the preceding section.  In particular for
  $\tonetwo$ hopping we explicitly evaluate the DOS and also the local
  Green function. The case of exponentially decreasing hopping is also
  discussed.  Furthermore we show how to construct a tight-binding
  Hamiltonian from a given DOS.

  \subsection{Dispersion relations for tight-binding Hamiltonians}\label{subsec:disp}
  
  Since the hopping Hamiltonian $H_d$ on the Bethe lattice is a
  polynomial in $\tilde{H}_1$ [see Eq.~(\ref{eq:chebychev1})], its
  eigenfunctions are the same as those of $\tilde{H}_1$, and its
  eigenvalues can be expressed as a function of the eigenvalues of
  $\tilde{H}_1$.  Explicitly, for a general tight-binding Hamiltonian
  of the form~(\ref{eq:H}), i.e.,
  \begin{align}
    H
    &=
    \sum_{d=0}^\infty
    t_dH_d
    =
    \sum_{d=0}^\infty
    t_d^{\ast}\tilde{H}_d
    \,,\label{eq:H-scaled}
  \end{align}
  we find from Eq.~(\ref{eq:chebychev1}) that $H$ $=$
  $\calF(\tilde{H}_1)$ with
  \begin{align}
    \calF(x)
    &=
    \sum_{d=0}^\infty
    \Big(t_d^{\ast}-\frac{t_{d+2}^{\ast}}{K}\Big)
    U_d(x/2)
    \,,\label{eq:Fcal}
  \end{align}
  where we used again the scaling $t_d$ $=$ $t_d^{\ast}/K^{d/2}$ and
  $\tilde{H}_d$ $=$ $H_d/K^{d/2}$ to facilitate the discussion of the
  limit $Z$ $\to$ $\infty$.  The eigenvalues $\epsilon$ of $H$
  are thus related to the eigenvalues $\lambda$ of $\tilde{H}_1$ by
  the ``dispersion relation''
  \begin{align}
    \epsilon(\lambda)
    &=
    \calF(\lambda)
    \,,\label{eq:dispersion}
  \end{align}
  which provides a surprising analogy to tight-binding dispersions for
  crystal lattices with translational symmetries, with $\lambda$
  playing the role of crystal momentum.  As a consequence of
  Eq.~(\ref{eq:dispersion}) the DOS for $H$ can be
  obtained by a simple change of variables from that of $\tilde{H}_1$
  [see Eq.~(\ref{eq:rho1})],
  \begin{align}
    \rho(\epsilon)
    &=
    \int\limits_{-2}^{2}
    \rho_1(\lambda)
    \,
    \delta(\epsilon-\epsilon(\lambda))
    \,
    d\lambda
    \label{eq:rho-delta}\\
    &=
    \sum_i
    |\lambda_i'(\epsilon)|
    \,\rho_1(\lambda_i(\epsilon)),
    \label{eq:rho-trafo}
  \end{align}
  where the sum runs over all solutions $\lambda_i(\epsilon)$ of
  $\epsilon=\epsilon(\lambda)$. Explicit eigenfunctions
  $\ket{\theta}$ and eigenvalues $\lambda(\theta)$ of $\tilde{H}_1$
  were obtained by Mahan in Ref.~\onlinecite{mahan}.
  
  We see that using Eqs.~(\ref{eq:H-scaled})-(\ref{eq:rho-trafo}) the
  DOS can be easily obtained for any tight-binding Hamiltonian.  At
  most a simple numerical inversion of the polynomial
  $\epsilon(\lambda)$ needs to be performed.  Below we discuss the
  case of $\tonetwo$ hopping as well as an exemplary case of
  long-range hopping, for which this inversion can be performed
  analytically.

  \subsection{$\tonetwo$ hopping}\label{subsec:onetwodos}
  
  The DOS for the Hamiltonian with nearest-neighbor ($\tone$) and
  next-nearest-neighbor ($\ttwo$) hopping can be found from
  \begin{align} 
    H_{\txone ,\txtwo}
    &:=
    \tone H_1+\ttwo H_2
    =
    \txone\tilde{H}_1+\txtwo\tilde{H}_2
    \nonumber\\
    &=
    \txtwo\tilde{H}_1^2+\txone\tilde{H}_1-p\txtwo\openone
    \,,\label{eq:H-ttprime}
  \end{align}
  where again $p$ $=$ $Z/K$, and $\tone$ $=$ $\txone/\sqrt{K}$ and
  $\ttwo$ $=$ $\txtwo/K$ were scaled appropriately, such that the
  limit $Z$ $\to$ $\infty$ can be taken.  The dispersion relation and
  the roots are given by
  \begin{align}
    \epsilon(\lambda)
    &=
    \txtwo\lambda^2+\txone\lambda-\txtwo p
    \,,\label{eq:eps-ttprime}
    \\
    \lambda_{1,2}(\epsilon)
    &=
    \frac{-\txone\pm\sqrt{\txonesq 
        +4\txtwo (p\txtwo +\epsilon)}}{2\txtwo}
    \,.\label{eq:lambda-ttprime}
  \end{align}
  This yields the DOS
  \begin{align}
    \rho_{\txone ,\txtwo}(\epsilon)
    &=
    \frac{1}{2\pi}
    \frac{
      \Theta(\txonesq +4\txtwo (p\txtwo +\epsilon))
    }{
      \sqrt{\txonesq +4\txtwo (p\txtwo +\epsilon)}
    }
    \sum_{i=1}^{2}
    \frac{\sqrt{4-\lambda_{i}(\epsilon)^2}}{p-\lambda_{i}(\epsilon)^2/Z}
    \,,\label{eq:rho-ttprime}
    \\
    &=
    \rho_{-\txone ,\txtwo}(\epsilon)
    =
    \rho_{\txone ,-\txtwo}(-\epsilon)
    \,,\label{eq:rho-ttprime-symmetry}
  \end{align}
  where $\Theta(x)$ denotes the step function.  In the limit $Z$ $\to$
  $\infty$ this simplifies to
   \begin{align}
    \rho^{\infty}_{\txone ,\txtwo}(\epsilon)
    &=
    \frac{
      \Theta(\txonesq +4\txtwo (\txtwo +\epsilon))
    }{
      \sqrt{\txonesq +4\txtwo (\txtwo +\epsilon)}
    }
    \sum_{i=1}^{2}
    \frac{
      \sqrt{4-\lambda^{\infty}_{i}(\epsilon)^2}
    }{2\pi}
    \,,\label{eq:rho-ttprime-infZ}
    \\
    \lambda^{\infty}_{1,2}(\epsilon)
    &=
    \frac{-\txone\pm\sqrt{\txonesq 
        +4\txtwo (\txtwo +\epsilon)}}{2\txtwo}
    \,.\label{eq:lambda-ttprime-infZ}
  \end{align}
  Note that for $\txtwo\neq0$ the DOS is asymmetric,\cite{blumer} in
  contrast to the case of \emph{random} $\tonetwo$ hopping on the Bethe
  lattice.\cite{rozenberg2,georges}

\begin{figure}[t]
    \centerline{\includegraphics[clip,width=\hsize,angle=0]{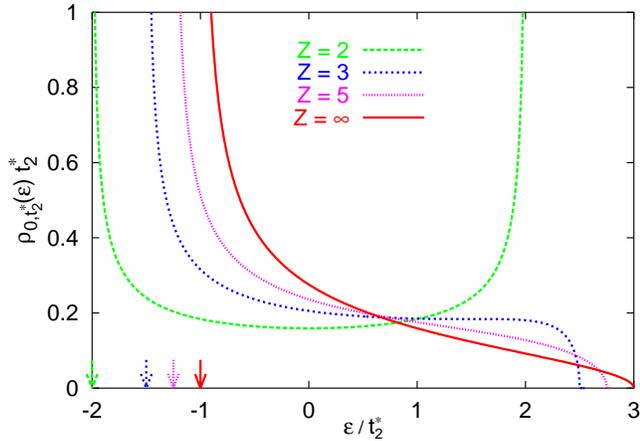}}
    \nopagebreak
    \caption{Density of states (\ref{eq:rho-tprime}) for pure NNN hopping 
      $\ttwo=\txtwo/(Z-1)\neq 0$ on the Bethe lattice, for
      several numbers of nearest neighbors $Z$;  $\txtwo =1$ sets the
      energy scale. Divergences are marked on the horizontal axis.}
    \label{fig:dos1}
  \end{figure}

  We now discuss some limiting cases of Eqs.~(\ref{eq:rho-ttprime}) and
  (\ref{eq:rho-ttprime-infZ}).  For $\txtwo$ $\to$ $0$ (pure NN
  hopping) we recover Eq.~(\ref{eq:rho1}) since then one of the roots 
  $\lambda_i$ reduces to $\epsilon/\txone $ while the other diverges.
  On the other hand, for $\txone $ $\to$ $0$ the DOS for
  pure NNN hopping (with $|\txtwo|$ $=$ $1$) is obtained 
  \begin{align}
    \rho_{0,1}(\epsilon)
    &=
    \frac{1}{2\pi}
    \frac{\sqrt{4-p-\epsilon}}{(1-\epsilon/Z)\sqrt{p+\epsilon}}
    =
    \rho_{0,-1}(-\epsilon) 
    \,.\label{eq:rho-tprime}    
  \end{align}
  In the limit $Z$ $\to$ $\infty$ this reduces to\cite{blumer}
  \begin{align}
    \rho^{\infty}_{0,1}(\epsilon)
    &= 
    \frac{1}{2\pi}\frac{\sqrt{3-\epsilon}}{\sqrt{1+\epsilon}}
    =
    \rho^{\infty}_{0,-1}(-\epsilon) 
    \,.\label{eq:rho-tprime-infZ}
  \end{align}
  In Fig.~\ref{fig:dos1} $\rho_{0,\txtwo}(\epsilon)$ (pure NNN
  hopping) is plotted for several values of $Z$.  This function has a
  square-root singularity on the left-hand side of the band.  In the case
  $Z=2$ the DOS has the same form as for a chain with NN hopping,
  because then the Bethe lattice with NNN hopping reduces to two
  disconnected infinite chains.  The DOS for $\tonetwo$ hopping on the
  triangular Husimi cactus is obtained in \husimianhang.

\begin{figure}[t]
    \centerline{\includegraphics[clip,width=\hsize,angle=0]{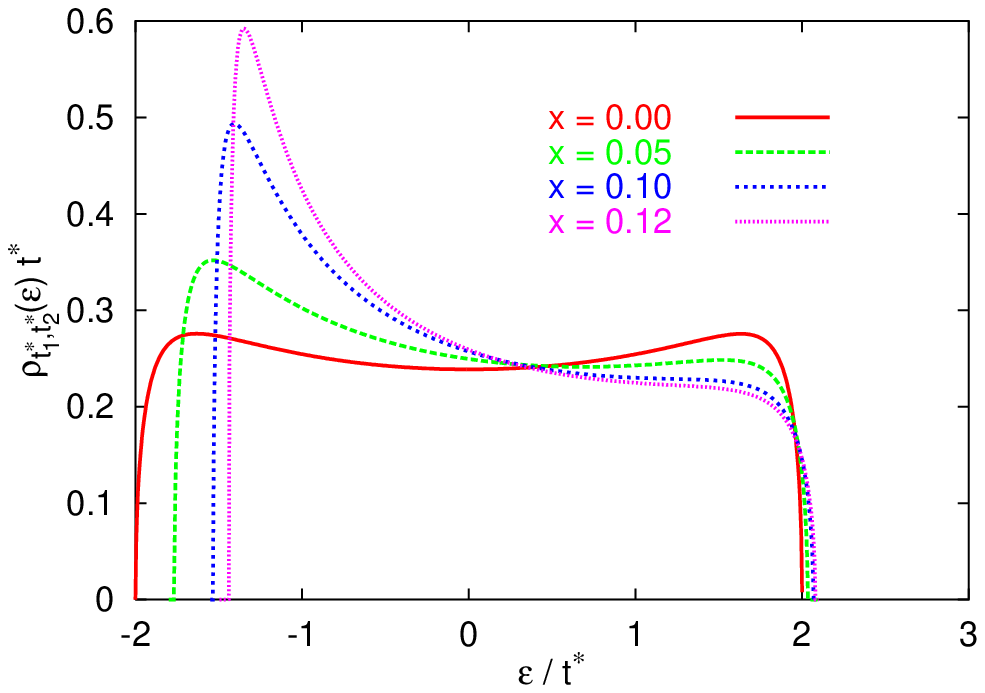}}
    \nopagebreak
    \centerline{\includegraphics[clip,width=\hsize,angle=0]{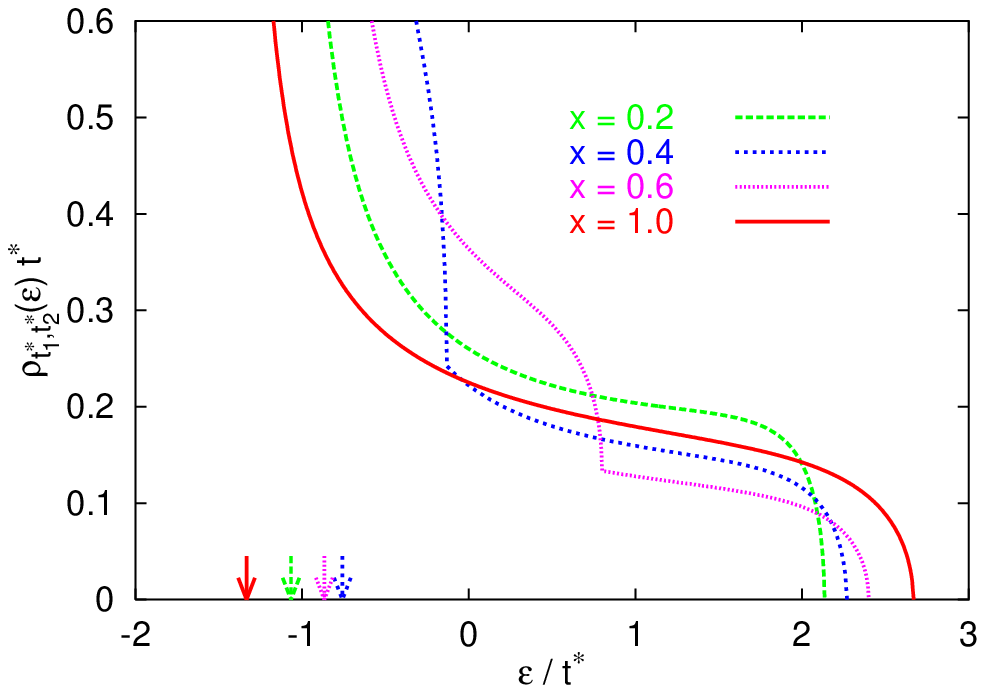}}
    \nopagebreak
    \caption{Density of states 
      (\ref{eq:rho-ttprime}) for $\tonetwo$ hopping on the Bethe
      lattice with $Z=4$ for selected values of $x$ $=$
      $\txtwo/\txpre$, where $\txone, \txtwo$ $\geq$ $0$, $\txpre$ $=$
      $\txone+\txtwo$ sets the energy scale.  Divergences are marked
      on the horizontal axis.}
    \label{fig:nnn1}
  \end{figure}

\begin{figure}[t]
    \centerline{\includegraphics[clip,width=\hsize,angle=0]{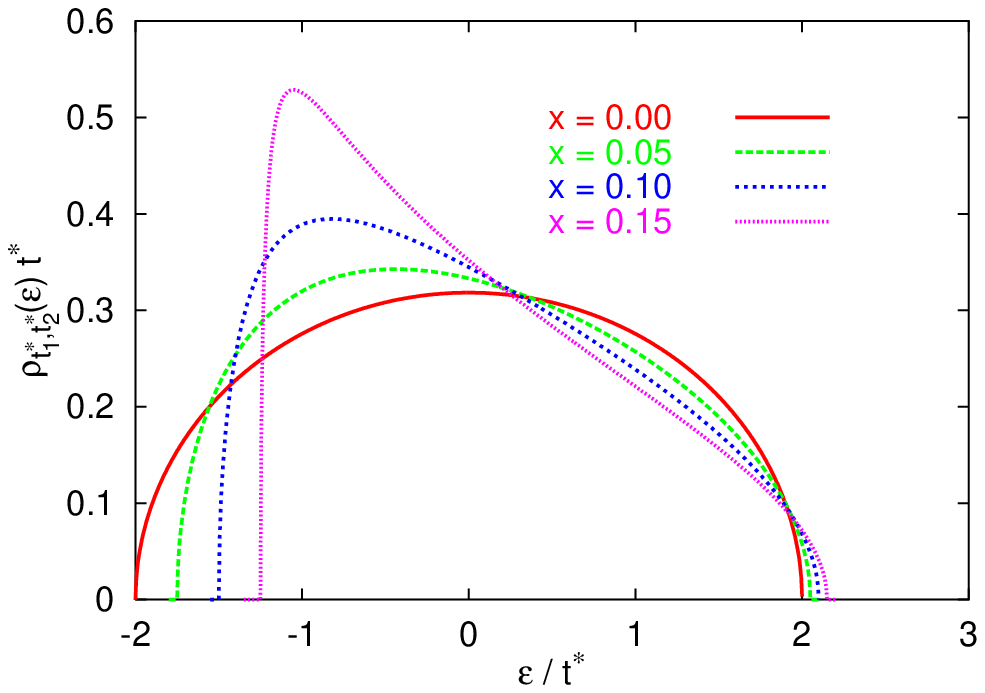}}
    \nopagebreak
    \centerline{\includegraphics[clip,width=\hsize,angle=0]{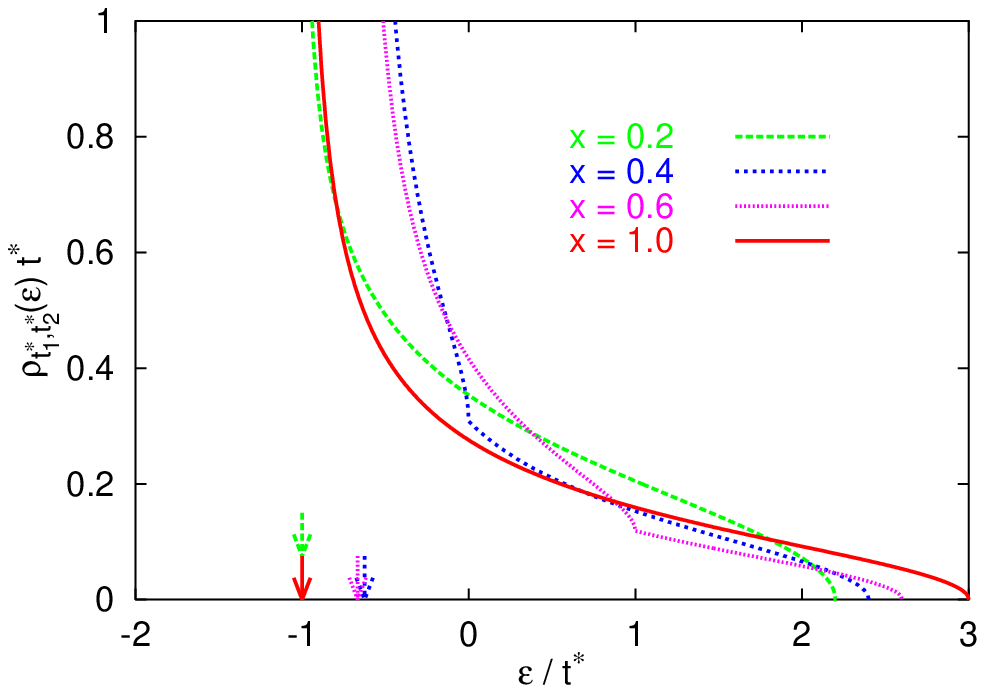}}
    \nopagebreak
    \caption{Same as Fig.~\ref{fig:nnn1}, but for $Z$ $\to$ $\infty$
      [Eq.~(\ref{eq:rho-ttprime-infZ})].}
    \label{fig:nnn2}
  \end{figure}

  In general, the shape of $\rho_{\txone ,\txtwo}(\epsilon)$ is
  determined by two dimensionless parameters, the coordination number
  $Z$ and the ratio $\txtwo/\txone$. In view of
  Eq.~(\ref{eq:rho-ttprime-symmetry}) it is sufficient to consider
  $\txone, \txtwo$ $\geq$ $0$. In the following we use the parameter
  $x$ $=$ $\txtwo/(\txone +\txtwo)$ $\in$ $[0,1]$, i.e., $\txtwo
  =x(\txone+\txtwo) $, $\txone =(1-x)(\txone+\txtwo)$.  In
  Figs.~\ref{fig:nnn1} and \ref{fig:nnn2} we present plots of
  $\rho_{\txone ,\txtwo}(\epsilon)$ for several values of the
  parameter $x$ in the case $Z$ $=$ $4$ and in the limiting case $Z$
  $\to$ $\infty$.  At small $x$ ($\txone\gg\txtwo$) the asymmetry of
  the DOS develops gradually.  At some critical value $x_*$ $=$ $1/5$
  a square-root singularity appears at the left-hand side of the band. For
  $x$ $<$ $x_*$ the DOS is a smooth function vanishing at both band
  edges, whereas for $x$ $>$ $x_*$ there is a singularity at the lower
  band edge as well as a cusp within the band with diverging first
  derivative $d\rho_{\txone ,\txtwo}(\epsilon)/d \epsilon$.  With
  increasing $x$ the cusp moves continuously to the upper band edge
  where it disappears for $x$ $=$ $1$.  These features can be
  understood with the help of the dispersion relation
  Eq.~(\ref{eq:eps-ttprime}).  The solutions $\lambda_{1,2}$
  [Eq.~(\ref{eq:lambda-ttprime})] of Eq.~(\ref{eq:eps-ttprime})
  contribute only if they lie in the interval $[-2,2]$ of nonvanishing
  $\rho_1(\lambda)$.  While for $x$ $<$ $x_*$ only $\lambda_1$
  contributes, for $x$ $>$ $x_*$ both solutions contribute if
  $\epsilon$ lies between the left band edge and the cusp, whereas
  only one solution contributes if $\epsilon$ lies above the cusp.

  Finally we note that the local Green function can be obtained from
  the DOS by a Kramers-Kronig relation, i.e.,
  \begin{align}
    G^{\infty}_{\txone ,\txtwo}(z)
    &=
    \int_{-\infty}^{\infty}
    \frac{\rho_{\txone ,\txtwo}(\epsilon)}{z-\epsilon}
    d\epsilon
    \,.\label{eq:kramerskronig}
  \end{align}
  For the case $Z$ $\to$ $\infty$ this leads to the result
   \begin{align}
    G^{\infty}_{\txone ,\txtwo}(z)
    &=
    \bra{i}
    (z-\txone \tilde{H}_1-\txtwo \tilde{H}_2)^{-1}
    \ket{i}
    \\
    &=
    \txtwo
    \frac{
      G^{\infty}_{\txtwo,0}(z_1)-G^{\infty}_{\txtwo,0}(z_2)
    }{
      z_1-z_2
    }
    \,,\label{eq:G-ttprime-infZ}
  \end{align}
  where $z_{1,2}$ $=$
  $(-\txone\pm\sqrt{\txonesq+4\txtwo(z+\txtwo)})/2$ and
  $G^{\infty}_{\txpre,0}(z)$ $=$
  $(z-\sqrt{z^2-4\txpresq})/(2\txpresq)$ is the usual local Green
  function for pure NN hopping as obtained from
  (\ref{eq:G1-nonlocal}). For pure $\ttwo$ hopping
  (\ref{eq:G-ttprime-infZ}) reduces to
  \begin{align}
    G^{\infty}_{0,\txtwo}(z)
    &=
    \frac{1}{2\txtwo}
    -
    \frac{1}{\txtwo}
    \sqrt{
      \frac{1}{4}
      -
      \frac{1}{1 + z/\txtwo}
    }
    \,,\label{eq:G-tprime-infZ}
  \end{align}
  with the square root given by its principal branch.
  The density of states and Green function for pure NNN hopping
  [Eqs.~(\ref{eq:rho-tprime-infZ}) and (\ref{eq:G-tprime-infZ})] were
  previously obtained in Ref.~\onlinecite{blumer} using RPE.

  \subsection{Long-range hopping}
  
\begin{figure}[t]
    \centerline{\includegraphics[clip,width=\hsize,angle=0]{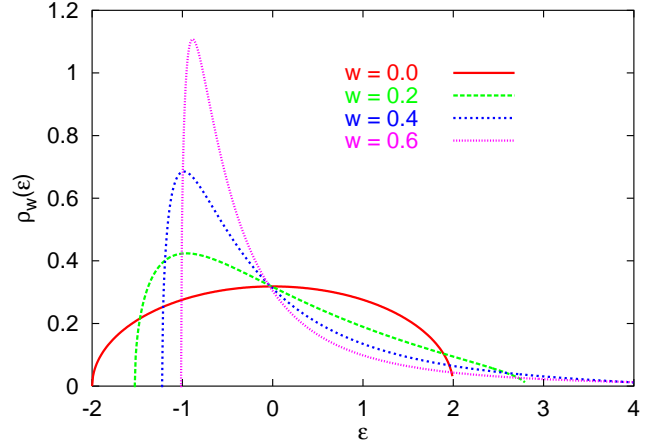}}
    \nopagebreak
    \caption{Density of states for exponentially 
      decreasing long-range hopping on the Bethe lattice
      [Eq.~(\ref{eq:H-w})], for $\txpre$ $=$ $1$ and $Z$ $\to$
      $\infty$.}
    \label{fig:rho-w}
  \end{figure}

  For long-range hopping beyond NNN the expressions for
  $\lambda(\epsilon)$ and $\rho(\epsilon)$ become rather complicated.
  However, the special case of exponentially decreasing hopping amplitudes,
  $t_d=w^{d-1}\txpre$, with Hamiltonian
  \begin{align}
    H_w
    =
    \txpre
    \sum_{d=1}^{\infty}
    w^{d-1}\tilde{H}_d
    \,,\label{eq:H-w}
  \end{align}
  allows for an analytical solution (here we assume $|w|<1$ to ensure
  convergence).  The familiar case of pure $\tone$ hopping corresponds
  to $w$ $\to$ $0$.  From Eq.~(\ref{eq:geometricseries}) it is
  straightforward to obtain the corresponding dispersion relation as
  \begin{align}
    \epsilon_w(\lambda)
    &=
    \frac{\txpre}{w}\left(\frac{1-w^2/K}{1-w\lambda + w^2}-1\right)
    \,,\label{eq:dispersion-w}
  \end{align}
  i.e., $\epsilon_w(\lambda)$ $=$ $\txpre\lambda$ $+$ $O(w)$.  In the
  interval $-2\le \lambda \le 2$, $\epsilon_w(\lambda)$ is a
  monotonous function with inverse function $\lambda_w(\epsilon)$. The
  DOS $\rho(\epsilon)$ is again calculated using
  Eq.~(\ref{eq:rho-trafo}),
  \begin{align}
    \rho_w(\epsilon)
    &=
    \frac{\txpre}{2\pi}
    \frac{1-w^2/K}{(w\epsilon+\txpre)^2}
    \frac{\sqrt{4-\lambda_w(\epsilon)^2}}{p-\lambda_w(\epsilon)^2/Z}
    \,,
  \end{align}
  where again $p$ $=$ $Z/K$ $=$ $1+1/K$.  In Fig.~\ref{fig:rho-w}
  $\rho_w(\epsilon)$ is plotted for several values of $w$ in the limit
  $Z$ $\to$ $\infty$.  For $|w|$ $<$ $1$ we find a smooth DOS 
  with finite bandwidth, while for $w\to1$ the upper band edge
  (where $\lambda_w(\epsilon)=2$) moves to infinity and the DOS
  decreases like $\epsilon^{-5/2}$ for large $\epsilon$.
   
\begin{figure}[t]
    \centerline{\includegraphics[clip,width=\hsize,angle=0]{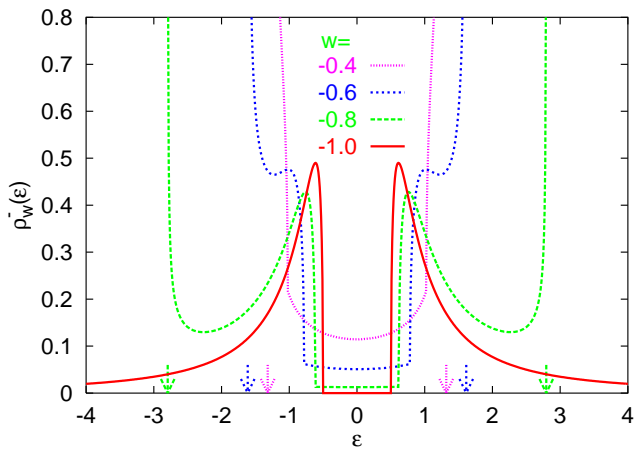}}
    \nopagebreak
    \caption{Density of states for exponentially 
      decreasing long-range hopping between different sublattices
      (``odd hopping'') for $\txpre$ $=$ $1$, $Z$ $\to$ $\infty$, and
      $w$ $<$ $w_*$. Divergences are marked on the horizontal axis.}
    \label{fig:rho-w+}
  \end{figure}

\begin{figure}[t]
    \centerline{\includegraphics[clip,width=\hsize,angle=0]{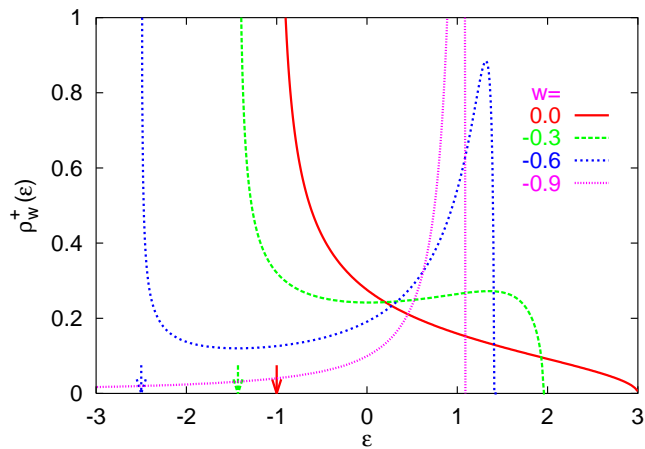}}
    \nopagebreak
    \caption{Density of states for exponentially 
      decreasing long-range hopping between same sublattices (``even
      hopping'') for $\txpre$ $=$ $1$, $Z$ $\to$ $\infty$, and $w$
      $\leq$ $0$. Divergences are marked on the horizontal axis.
      For $w$ $=$ -0.9 the finite peak at the upper band
      edge, as well as the singularity at the lower band edge, are
      outside the plotting range.}
    \label{fig:rho-w-}
  \end{figure}

  It is also interesting to consider only ``odd'' hopping, i.e.,
  between different A and B sublattices, or ``even'' hopping between
  the same sublattices, although there is no immediate physical
  motivation for this restriction.  The Hamiltonians $H_w^-$ $=$
  $\txpre \sum_{d=1}^{\infty} w^{d-1}\tilde{H}_{2d-1}$ and $H_w^+$ $=$
  $\txpre \sum_{d=1}^{\infty} w^{d-1}\tilde{H}_{2d}$ describe odd and
  even hopping with exponentially decreasing amplitudes, respectively.
  They lead to the dispersion relations
  \begin{align}
    \epsilon_w^-(\lambda)
    &=
    \txpre\frac{(1-w/K)\lambda}{(1+w)^2-w\lambda^2}
    \,,
    \\
    \epsilon_w^+(\lambda)
    &=
    \frac{\txpre}{w}
    \left(
      \frac{(1+w)(1-w/K)}{(1+w)^2-w\lambda^2}
      -
      1
    \right)
    \,,
  \end{align}
  which are antisymmetric and symmetric in $\lambda$, respectively,
  and yield finite bandwidths except for $w=1$.  Thus for odd hopping
  the DOS $\rho_w^-(\epsilon)$ is symmetric as in
  Fig.~\ref{fig:rho-w+}, whereas for even hopping $\rho_w^+(\epsilon)$
  is asymmetric (Fig.~\ref{fig:rho-w-}).  We note that
  $\rho_w^-(\epsilon)$ is finite for $w$ $>$ $w_*$ $:=$ $3-2\sqrt{2}$
  $\approx$ $-0.172$, while square-root singularities at the band
  edges occur for $w$ $<$ $w_*$.  For the latter case another
  remarkable feature is that the DOS is nearly constant in the middle
  of the band, up to a cusp where it rises sharply (see
  Fig.~\ref{fig:rho-w+}).  On the other hand, $\rho_w^+(\epsilon)$
  always has a square-root singularity at one band edge [$\epsilon$
  $=$ $-p\txpre/(1+w)$], similar to the case of pure NNN hopping.
  Note that for $w$ $\to$ $-1$ this singularity moves to infinity
  albeit with vanishing weight; most of the weight is near the other
  band edge in this limit.
  
  We note that by naively setting $w=\sqrt{K}$ (without addressing
  questions of convergence) in the above results one recovers the case
  of infinite-range hopping, where each site is connected to every
  other site (``complete graph'') and the underlying lattice structure
  becomes irrelevant.  The energy $\epsilon(\lambda)$ becomes constant
  in this case and we find
  $\rho_w(\epsilon)\to\delta(\epsilon+\txpre/\sqrt{K})$, in agreement
  with the result for infinite-range hopping on the hypercubic
  lattice, for which the corresponding Hubbard model is
  solvable.\cite{vandongen}

  \subsection{Tight-binding Hamiltonian for arbitrary density 
    of states}\label{subsec:arbdos}
  
  In mean-field theories the dependence on the lattice typically
  enters only through the DOS for free particles. For simplicity often
  DOSs with a particularly simple form are employed, but when
  calculating two-particle quantities or going beyond mean-field
  theory it is necessary to know the corresponding hopping amplitudes.
  They are also useful for the realistic modelling of strongly
  correlated materials where the DOS is obtained from \emph{ab initio}
  calculations.\cite{PT} In this section we show how to determine a
  tight-binding Hamiltonian $H$ that corresponds to a given (one-band)
  DOS $\rho(\epsilon)$. For hypercubic lattices this task has already
  been addressed in Refs.~\onlinecite{wahle98,blumer,blumer2,kollar}.
  Here we perform such a construction for the Bethe lattice, based on
  the relation (\ref{eq:geometricseries}).

  We start again from the general hopping Hamiltonian
  (\ref{eq:H-scaled}).  As noted above, the eigenvalues of $H$ are
  given by the dispersion relation $\epsilon(\lambda)$ $=$
  $\calF(\lambda)$ [see Eq.~(\ref{eq:Fcal})].  We can thus use the
  methods of Bl\"umer\cite{blumer,blumer2} who found a similar map
  between the spectra of $H$ and $\tilde{H}_1$ for the hypercubic
  lattice in the limit of high dimensions.  In the present case we
  choose
  \begin{align}
    \calF(\lambda)
    &=
    \mu(n_1(\lambda))
    \,,
  \end{align}
  where $n_1(\lambda)$ $=$
  $\int_{-2}^{\lambda}\rho_1(\lambda')\,d\lambda'$ and $n(\mu)$ $=$
  $\int_{\epsilon_{\text{min}}}^{\mu}\rho(\epsilon)\,d\epsilon$. Here
  $\epsilon_{\text{min}}$ is the lower band edge of $\rho(\epsilon)$
  and $\mu(n)$ denotes the inverse function of $n(\mu)$.  Then
  $\calF'(\lambda)$ $>$ $0$ for $-2$ $<$ $\lambda$ $<$ $2$, and the
  DOSs are related by
  \begin{align}
    \rho_1(\lambda)
    &=
    \calF'(\lambda)\,\rho(\calF(\lambda))
    \,,
  \end{align}
  which indeed implies Eq.~(\ref{eq:rho-delta}).  It remains to
  determine the hopping parameters in Eq.~(\ref{eq:H-scaled}) for this
  choice of $\calF(\lambda)$.  Using the orthogonality of the
  Chebyshev polynomials\cite{abramowitz84a} we find that
  $t_{d+2}^{\ast}/K$ $=$ $t_{d}^{\ast}-u_d$, where
  \begin{align}
    u_d
    &=
    \int\limits_{-2}^{2}
    \rho^{\infty}_1(\lambda)
    \calF(\lambda)
    U_d(\lambda/2)
    \,d\lambda
    \,.
  \end{align}
  
  For the remainder of this section we will consider only the limit
  $Z$ $\to$ $\infty$, for which we have $t_d^{\ast}$ $=$ $u_d$ and
  $n_1(\lambda)$ $=$
  $[1+\lambda\rho^{\infty}_1(\lambda)]/2+\arcsin(\lambda/2)/\pi$, and
  $u_0$ $=$ $\int\!\epsilon\rho(\epsilon)\,d\epsilon$ $=:$ $M_1$
  yields the first moment of the target DOS.  Even then the
  calculation of the hopping amplitudes will typically involve
  numerical integrations.  As an analytically tractable example we
  consider the model DOS
  \begin{align}
    \rho(\epsilon)
    &=
    c
    \frac{\Theta(D-|\epsilon|)}{\sqrt{1+a\epsilon/D}}
    \,,\label{eq:dos-squarerootsing}
    &
    c
    &=
    \frac{\sqrt{1+a}+\sqrt{1-a}}{4D}
    \,,
  \end{align}
  where $D$ is the half-bandwidth and the parameter $-1$ $\leq$ $a$
  $\leq$ $1$ determines the asymmetry, e.g., the first moment is given
  by $M_1$ $=$ $-a/(24Dc^2)$.  A constant (rectangular) DOS
   is recovered for $a$ $=$ $0$, and a square-root singularity
  at one band edge is present for $|a|$ $=$ $1$.  Note that this model
  DOS does not vanish at the band edges, in contrast to the DOS
  proposed
  in Ref.~\onlinecite{wahle98}.  For
  the latter, however, an analytical calculation of the corresponding
  hopping amplitudes appears infeasible.
  
\begin{figure}[t]
    \centerline{\includegraphics[clip,width=\hsize,angle=0]{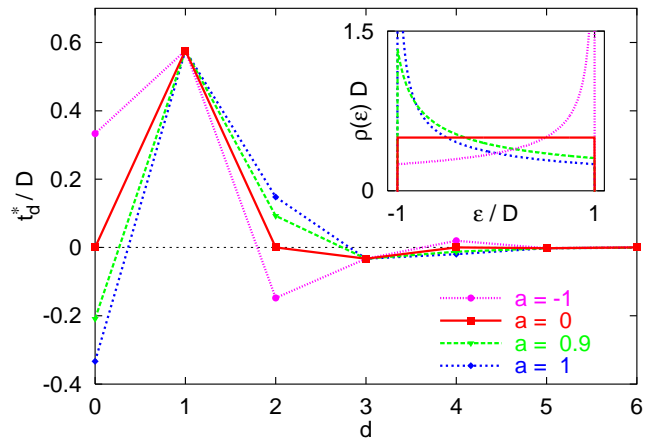}}
    \nopagebreak
    \caption{Hopping amplitudes $t_d^{\ast}$ on the Bethe lattice 
      (with $Z$ $\to$ $\infty$) corresponding to the DOS
      (\ref{eq:dos-squarerootsing}) (see inset) for several values of
      the asymmetry parameter $a$. The ``odd'' hopping amplitudes $t_1^*$,
      $t_3^*$, etc. are independent of $a$; $t_0^{\ast}$ gives the
      center of the mass of the band.}
    \label{fig:amplitudes}
  \end{figure}

  For the DOS (\ref{eq:dos-squarerootsing}) we have $\mu(n)$ $=$ $-D$
  + $n/\rho(-D)$ $-$ $6M_1n^2$. In general, upon expanding $\mu(n)$ as
  a power series in $n$, the calculation of $t_{d}^{\ast}$ involves
  the coefficients
  \begin{align}
    r_{d,s}
    &=
    \int\limits_{-2}^{2}
    \rho^{\infty}_1(\lambda)
    U_d(\lambda/2)
    [n_1(\lambda)]^s
    \,d\lambda
    \label{eq:rcoeff}
    \\
    &=
    \frac{2}{\pi}
    \int\limits_{0}^{\pi}
    \sin x
    \,
    \sin(d+1)x
    \,
    \bigg[\frac{2x-\sin2x}{2\pi}\bigg]^s
    dx
    \,,\label{eq:rcoeff-trig}
  \end{align}
  where we have used the representation\cite{abramowitz84a} $U_n(x)$
  $=$ $\sin[(n+1)\arccos x]/\sin(\arccos x)$ for the Che\-by\-shev
  polynomials. Evaluating Eq.~(\ref{eq:rcoeff-trig}) for $s$ $=$
  $0,1,2$ we obtain for the DOS (\ref{eq:dos-squarerootsing})
  \begin{subequations}%
  \begin{align}%
    t_{d}^{\ast}
    &=
    \left\{
      \begin{array}{ll}
        DR(d)
        &
        \text{for odd $d$}
        \\[1ex]
        -3M_1R(d)
        &
        \text{for even $d$}
      \end{array}
    \right.
    ,
    \\
    R(d)
    &=
    \left\{
      \begin{array}{ll}
        \tfrac{-1}{3}        &  \text{for $d$ $=$ $0$}\\[1ex]
        \tfrac{35}{24\pi^2}  &  \text{for $d$ $=$ $2$}\\[1ex]
        \tfrac{-7}{36\pi^2}  &  \text{for $d$ $=$ $4$}\\[1ex]
        \tfrac{-128(d+1)}{\pi^2(d-2)d^2(d+2)^2(d+4)}
                             &  \text{otherwise}
      \end{array}
    \right.
    .
  \end{align}%
  \end{subequations}%
  In Fig.~\ref{fig:amplitudes} we plot the hopping amplitudes
  $t_d^{\ast}$ vs $d$, which decay slower for large distances than
  in Eq.~(\ref{eq:H-w}), due to the algebraic behavior of $R(d)$
  $\sim$ $d^{-5}$.  Note that the amplitudes for hopping between the same
  sublattices (even $d$) are proportional to $M_1$ $=$ $O(a)$ and thus
  vanish for the special case $a$ $=$ $0$ (constant DOS). On the other
  hand, for different sublattices (odd $d$) the hopping amplitudes are
  independent of the asymmetry parameter $a$; except for $t_1^{\ast}$
  $\approx$ $0.58D$ they are all negative and drop off quickly (e.g.,
  $t_3^{\ast}$ $=$ $14t_5^{\ast}$ $\approx$ $-0.033D$).  Regarding
  $t_{d}^{\ast}/t_1^{\ast}$ the {\em quantitative} difference between
  hopping for constant and semielliptic DOS (pure NN
  hopping) is thus surprisingly small.

  \section{Dynamical mean-field theory }\label{sec:dmft}
  
  We now turn to dynamical mean-field theory
  (DMFT)\cite{Georges92,Jarrell92,vollha93,pruschke,georges,PT} for
  the Hubbard model, for which the Hamiltonian reads
  \begin{align}
    H_{\text{H}}
    &=
    H_{\text{hop}}
    +
    H_{\text{int}}
    +
    H_{\text{ext}}
    \,,\label{eq:Hubbard}
    \\
    H_{\text{hop}}
    &=
    \sum_{ij\sigma}
    t_{ij}
    c_{i\sigma}^{\dagger}
    c_{j\sigma}^{{\phantom{\dagger}}}
    \,,\label{eq:Hubbard-hop}
    \\
    H_{\text{int}}
    &=
    U
    \sum_{i}
    n_{i\uparrow}n_{i\downarrow}
    \,,\label{eq:Hubbard-int}
  \end{align}
  where $c_{i\sigma}$ and $c_{i\sigma}^{\dagger}$ are the usual
  fermionic annihilation and creation operators for site $i$ and spin
  $\sigma$, $n_{i\sigma}$ $=$
  $c_{i\sigma}^{\dagger}c_{i\sigma}^{{\phantom{\dagger}}}$ is number
  operator, and $H_{\text{ext}}$ involves external fields and is given below
  [Eq.~(\ref{eq:Hubbard-ext})].  DMFT becomes exact for this model in
  the limit of infinite coordination number, $Z$ $\to$ $\infty$.\cite{vollhardt}
  Below we will evaluate
  the DMFT equations for the Bethe lattice with arbitrary (in particular
  $\tonetwo$) hopping.

  \subsection{DMFT for the Hubbard model}
  
  We begin with a very brief summary of DMFT for the Hubbard model
  (see Ref.~\onlinecite{georges} for a review).  Let $\bm{G}$ and
  $\bm{\Sigma}$ denote the imaginary-time-ordered Green function and
  self-energy, which are matrices in site labels $i$, spin indices
  $\sigma$, and imaginary-time slices $\tau$ (or Matsubara frequencies
  $\mathrm{i}\omega_n$). They satisfy the Dyson equation
  \begin{align}
    \bm{G}
    &=
    [(\bm{G}^{(0)})^{-1}-\bm{\Sigma}]^{-1}
    \,,\label{eq:Dyson}
  \end{align}
  where $\bm{G}^{(0)}$ is the Green function for $U$ $=$ $0$. Due to
  the appropriate scaling of $t_{ij}$ the self-energy becomes
  \emph{local} in the limit $Z$ $\to$ $\infty$,
  \begin{align}
    (\bm{\Sigma})_{ij,\sigma,n}
    &=
    \Sigma_{i\sigma n}\delta_{ij}
    \,,\label{eq:selfenergy}
  \end{align}
  and its skeleton expansion depends only on the local Green function
  \begin{align}
    G_{i\sigma n}
    =
    (\bm G)_{ii,\sigma,n}
    =
    ([(\bm{G}^{(0)})^{-1}-\bm{\Sigma}]^{-1})_{ii,\sigma,n}
    \,.\label{eq:selfconsistency}
  \end{align}
  Therefore $G_{i\sigma n}$ and $\Sigma_{i\sigma n}$ can also be
  calculated from an auxiliary impurity problem\cite{Georges92} with
  the action
  \begin{multline}
    \calA_i
    =
    \sum_{n,\sigma}
    c_{i\sigma}^{\star}(\mathrm{i}\omega_n)
    \calG_{i\sigma n}^{-1}
    c_{i\sigma}^{\phantom{\star}}(\mathrm{i}\omega_n)
    \\
    -
    U
    \int_{0}^{\beta}
    c_{i\uparrow}^{\star}(\tau)
    c_{i\uparrow}^{\phantom{\star}}(\tau)
    c_{i\downarrow}^{\star}(\tau )
    c_{i\downarrow}^{\phantom{\star}}(\tau)
    \,d\tau
    \,,\label{eq:action}
  \end{multline}
  according to
  \begin{align}
    G_{i\sigma n}
    &=
    \langle
    c_{i\sigma}^{\star}(\mathrm{i}\omega_n)
    c_{i\sigma}^{\phantom{\star}}(\mathrm{i}\omega_n)
    \rangle_{\calA_i}
    \label{eq:G-from-A}
    \\
    &=
    [
    \calG_{i\sigma n}^{-1}
    -
    \Sigma_{i\sigma n}
    ]^{-1}
    \,,\label{eq:Dyson-imp}
  \end{align}
  so that the so-called Weiss field $\calG_{i\sigma n}$ can be
  eliminated and two equations, (\ref{eq:selfconsistency}) and
  (\ref{eq:G-from-A}), remain for $G_{i\sigma n}$ and $\Sigma_{i\sigma
    n}$.
  
  The DMFT equations (\ref{eq:selfconsistency})-(\ref{eq:Dyson-imp})
  are thus a closed set of equations for the self-energy of the
  Hubbard model (\ref{eq:Hubbard}), and become exact in the limit $Z$
  $\to$ $\infty$. In practice these equations are solved by iteration,
  and the dynamic impurity problem
  (\ref{eq:action})-(\ref{eq:Dyson-imp}) is solved approximately by
  numerical or diagrammatic methods.  In addition one must evaluate
  the self-consistency equation (\ref{eq:selfconsistency}), i.e.,
  $G_{i\sigma n}$ must be expressed in terms of $\Sigma_{i\sigma n}$
  and the noninteracting spectrum, preferably involving only the
  DOS. It is also useful to obtain $\calG_{i\sigma n}$
  in terms of $G_{i\sigma n}$ for use in Eq.~(\ref{eq:action}).  Below
  these evaluations are performed for $\tonetwo$ hopping on the Bethe
  lattice with $Z$ $\to$ $\infty$ for homogeneous phases and phases
  with broken sublattice symmetry.  For homogeneous phases
  $\Sigma_{i\sigma n}$ and $G_{i\sigma n}$ are independent of $i$,
  whereas for broken sublattice symmetry on a bipartite lattice one
  has $\Sigma_{i\sigma n}$ $=$ $\Sigma_{\gamma\sigma n}$ and
  $G_{i\sigma n}$ $=$ $G_{\gamma\sigma n}$ where $\gamma$ $=$ $(-1)^i$
  $=$ $\pm1$ $=$ A,B depending on the sublattice. We add homogeneous
  and staggered magnetic fields to the Hamiltonian,
  \begin{align}
    H_{\text{ext}}
    &=
    -
    \sum_{i\sigma}
    (h_{\text{f}}+h_{\text{af}}(-1)^i)
    \sigma n_{i\sigma}
    \,,\label{eq:Hubbard-ext}
  \end{align}
  i.e., with a local field $h_{i\sigma}$ $=$ $(h_{\text{f}}+(-1)^i
  h_{\text{af}})\sigma$, which allows one to detect ferromagnetic or
  antiferromagnetic phases.

  \subsection{Self-consistency equations}\label{subsec:selfconsistency}
  
  We now evaluate the self-consistency equation
  (\ref{eq:selfconsistency}) for arbitrary (scaled) hopping $t_{ij}$
  $=$ $t_{d_{ij}}^*/K^{d_{ij}/2}$ on the Bethe lattice. The effective
  dispersion is then $\epsilon(\lambda)$ $=$ $\calF(\lambda)$ with DOS
  $\rho(\epsilon)$, see Eqs.~(\ref{eq:Fcal})-(\ref{eq:rho-trafo}).  In
  particular we consider $\tonetwo$ hopping in the limit $Z$ $\to$
  $\infty$ with DOS (\ref{eq:rho-ttprime-infZ}).
  
  \subsubsection{Homogeneous phases} \label{subsubsec:hom}
  
  In the homogeneous case all sites are equivalent, as described
  above. We thus set $h_{\text{af}}$ $=$ $0$ but keep $h_{\text{f}}$,
  which possibly leads to a ferromagnetic response.  The
  self-consistency equation (\ref{eq:selfconsistency}) becomes
  \begin{align}
    G_{\sigma n}
    &=
    \int
    \frac{\rho(\epsilon)}{z_{\sigma n}-\epsilon}
    \,d\epsilon
    \label{eq:hilbert}
    \\&=
    \int\limits_{-2}^{2}
    \frac{\rho_1(\lambda)}{z_{\sigma n}-\calF(\lambda)}
    \,d\lambda
    \,,\label{eq:hilbert-lambda}
  \end{align}
  where $z_{\sigma n}=\mathrm{i}\omega_n+\mu+\sigma
  h_{\text{f}}-\Sigma_{n\sigma}$, $\mu$ is the chemical potential, and
  the dispersion relation $\calF(\lambda)$ [see
  Eqs.~(\ref{eq:Fcal})-(\ref{eq:dispersion})] has been substituted.  The
  general result (\ref{eq:hilbert-lambda}) thus yields the DMFT
  self-consistency equation in the homogeneous case \emph{for arbitrary
    hopping}.  The special case of $\tonetwo$ hopping is discussed in
  Sec.~\ref{subsec:weiss}.

  \subsubsection{Phases with broken sublattice symmetry}
  
  In the case of broken sublattice symmetry the local Green function
  and self-energy depend on $i$ through the sublattice index $\gamma$.
  Spontaneous breaking of this symmetry can be detected through the
  staggered magnetic field $h_{\text{af}}$; We also keep the
  homogeneous field $h_{\text{f}}$ which permits phases with
  $\Sigma_{A\sigma n}$ $\neq$ $\Sigma_{B\bar{\sigma} n}$.
  
  We start from the eigenbasis of the NN hopping
  Hamiltonian $\tilde{H}_1$, i.e., $\tilde{H}_1\ket{\theta}$ $=$
  $\lambda(\theta)\ket{\theta}$; these eigenstates may later be
  identified with those of Ref.~\onlinecite{mahan}. We use the
  sublattice transformation $\ket{\bar{\theta}}$ $=$
  $\sum_i(-1)^i\ket{i}\braket{i}{\theta}$, which yields
  $\tilde{H}_1\ket{\bar{\theta}}$ $=$
  $-\lambda(\theta)\ket{\bar{\theta}}$, and introduce wave functions
  that have nonzero amplitudes only on sublattice $\gamma$, i.e.,
  $\ket{\gamma\theta}$ $=$ $(\ket{\theta}+\gamma
  \ket{\bar{\theta}})/\sqrt{2}$.  Using the corresponding fermion
  operators $c_{\gamma\theta\sigma}^{\phantom{\dagger}}$ we now
  transform the tight-binding part $H_{\text{hop}}$ of the Hubbard
  model with effective dispersion $\calF(\lambda(\theta))$,
  \begin{multline}
    H_{\text{hop}}
    +
    H_{\text{ext}}
    =
    \sum_{\substack{\theta\sigma\\\lambda(\theta)>0}}
    \Big(
      \begin{array}{cc}
        c_{A\theta\sigma}^{\dagger}
        &
        c_{B\theta\sigma}^{\dagger}
      \end{array}
    \Big)
    \\
    \Bigg(
      \begin{array}{cc}
        \calF_+(\lambda(\theta))-h_{\text{A}\sigma}
        &
        \calF_-(\lambda(\theta))
        \\
        \calF_-(\lambda(\theta))
        &
        \calF_+(\lambda(\theta))-h_{\text{B}\sigma}
      \end{array}
    \Bigg)
    \Bigg(
      \begin{array}{c}
        c_{A\theta\sigma}^{\phantom{\dagger}}
        \\
        c_{B\theta\sigma}^{\phantom{\dagger}}
      \end{array}
    \Bigg)
  \end{multline}
  where $h_{\gamma\sigma}$ $=$ $(h_{\text{f}}+\gamma
  h_{\text{af}})\sigma$ and $\calF_{\pm}(\lambda)$ $=$
  $[\calF(\lambda)\pm\calF(-\lambda)]/2$. Note that the reduced
  interval for $\lambda(\theta)$ is analogous to the halving of the
  magnetic Brillouin zone for regular crystal lattices.
  
  For the evaluation of the self-consistency equation
  (\ref{eq:selfconsistency}) we include the self-energy and perform
  the matrix inversion. This yields the interacting local Green
  function as a function of the local self-energy
  \begin{align}
    G_{\gamma\sigma n}
    &=
    \sum_{\substack{\theta\\\lambda(\theta)>0}}
    \frac{
      |\braket{i}{\gamma\theta}|^2
      \;
      [z_{\bar{\gamma}\sigma n}-\calF_+(\lambda(\theta))]
    }{
      \prod_{\gamma'}
      [z_{\gamma'\sigma n}-\calF_+(\lambda(\theta))]
      -\calF_-(\lambda(\theta))^2
    }
    \nonumber
    \\
    &=
    \int\limits_{-2}^{2}
    \frac{
      \rho_1(\lambda)
      \;
      [z_{\bar{\gamma}\sigma n}-\calF_+(\lambda)]
    }{
      \prod_{\gamma'}
      [z_{\gamma'\sigma n}-\calF_+(\lambda)]
      -\calF_-(\lambda)^2
    }
    \,d\lambda
    \,,\label{eq:green-AB-rho}
  \end{align}
  where $i$ is any site belonging to sublattice $\gamma$ and
  $z_{\gamma\sigma n}=\mathrm{i}\omega_n
  +\mu+h_{\gamma\sigma}-\Sigma_{\gamma \sigma n}$.
  Equation (\ref{eq:green-AB-rho}) thus yields the DMFT self-consistency
  equations \emph{for arbitrary hopping}.
  
  The general expressions Eqs.~(\ref{eq:hilbert-lambda}) and
  (\ref{eq:green-AB-rho}) are the central results of our paper
  regarding DMFT.  Note that their derivation involved no counting of
  lattice paths or other combinatorial efforts. Rather they were made
  possible by the dispersion relation $\calF(\lambda)$
  [Eqs.~(\ref{eq:Fcal})-(\ref{eq:dispersion})] which is due to the
  operator identity (\ref{eq:geometricseries}).

  \subsection{$\tonetwo$ hopping}\label{subsec:weiss}
  
  We now specialize to $\tonetwo$ hopping [with dispersion relation
  (\ref{eq:eps-ttprime})] and the limit $Z$ $\to$ $\infty$.  For the
  homogeneous case we find from (\ref{eq:hilbert-lambda}) that
  $G_{\sigma n}$ $=$ $G^{\infty}_{\txone ,\txtwo}(z_{\sigma n})$,
  where the latter function was obtained in
  Sec.~\ref{subsec:onetwodos}
  [Eqs.~(\ref{eq:kramerskronig}-\ref{eq:G-tprime-infZ})]. The
  resulting Weiss field is discussed below
  [Eq.~(\ref{eq:weiss-ttprime-hom})].
  
  For broken sublattice symmetry we use Eq.~(\ref{eq:green-AB-rho})
  with $\calF_+(\lambda)$ $=$ $(\lambda^2-1)\txtwo $ and
  $\calF_-(\lambda)$ $=$ $\lambda \txone $.  Performing the integral
  in Eq.~(\ref{eq:green-AB-rho}) we find
  \begin{multline}
    G_{\gamma \sigma n}
    =
    \frac{1}{2\txtwo}
    +
  \\
    \sum_{i=1}^2
    \frac{
      (-1)^i
      [
      z_{\bar{\gamma}\sigma n}
      -
      (\lambda_i^2-1)\txtwo
      ]
      \sqrt{\lambda_i-2}\sqrt{\lambda_i+2}
    }{
      2
      (\lambda_2^2-\lambda_1^2)\lambda_i\txtwosq 
    },
    \label{eq:green-AB-ttprime}
  \end{multline}
  where $\pm\lambda_i$ are the poles of the integrand, given by
  \begin{subequations}%
    \begin{align}%
      \lambda_{i}
      &=
      \sqrt{\calA\pm\sqrt{\calA^2-\calB}}
      \,,
      \\
      \calA
      &=
      1+\frac{\txtwo (z_{A\sigma n}+z_{B\sigma n})+\txonesq }{2\txtwosq }
      \\
      \calB
      &=
      \bigg(\frac{z_{A\sigma
          n}}{\txtwo}+1\bigg)\bigg(\frac{z_{B\sigma n}}{\txtwo}+1\bigg)
      \,,
    \end{align}%
    \label{eq:lambda-AB}%
  \end{subequations}%
  and all square roots are given by their principal branches.
  A more compact expression results if one solves for $z_{\gamma\sigma
    n}$ in terms of Green functions, i.e.,
  \begin{multline}
    z_{\gamma\sigma n}
    =
    -\txtwo 
    \\
    +
    \frac{1}{G_{\gamma\sigma n}(1-\txtwo G_{\gamma\sigma n})}
    +
    \frac{
      \txonesq G_{\bar{\gamma}\sigma n}(1-\txtwo G_{\bar{\gamma}\sigma n})
    }{
      (1-\txtwo\sum_{\gamma'}G_{\gamma'\sigma n})^2
    }
    \,.\label{eq:z-AB}
  \end{multline}
    
  In the present notation the Weiss field is given by
  $\calG_{\gamma\sigma n}^{-1}$ $=$ $\mathrm{i}\omega_n + \mu +
  h_{\gamma\sigma} + G_{\gamma \sigma n}^{-1} - z_{\gamma \sigma n}$.
  From Eq.~(\ref{eq:z-AB}) follows the exact relation between Weiss field and 
  interacting local Green function as
  \begin{multline}
    \calG_{\gamma\sigma n}^{-1}
    =
    \mathrm{i}\omega_n
    +
    \mu
    +
    h_{\gamma\sigma}
    \\
    -
    \frac{
      \txonesq G_{\bar{\gamma}\sigma n}(1-\txtwo G_{\bar{\gamma}\sigma n})
    }{
      (1-\txtwo\sum_{\gamma'}G_{\gamma'\sigma n})^2
    }
    -
    \frac{\txtwosq G_{\gamma\sigma n}}{(1-\txtwo G_{\gamma\sigma n})}
    \,.\label{eq:weiss-ttprime-AB}
  \end{multline}
  Thus for nonrandom $\tonetwo$ hopping the self-consistency equation
  is more complicated than for the two-sublattice fully frustrated
  model with random hopping,\cite{georges,rozenberg2} in which 
  all terms in parentheses
  are absent.  Our algebraic derivation validates
  Ref.~\onlinecite{radke}, where Eq.~(\ref{eq:weiss-ttprime-AB}) was
  obtained using RPE methods, which required the classification of
  many complicated hopping processes.  We also note that for $U=0$
  Eq.~(\ref{eq:weiss-ttprime-AB}) reduces to the same quartic equation
  that follows from Ref.~\onlinecite{tanaskovic} for the
  noninteracting Green function [Eq.~(\ref{eq:G-ttprime-infZ})].

  The corresponding equation for the homogeneous case is obtained by
  setting $h_{\text{af}}$ to zero and dropping sublattice indices,
  i.e.,
  \begin{multline}
    \calG_{\sigma n}^{-1}
    =
    \mathrm{i}\omega_n
    +
    \mu
    +
    \sigma h_{\text{f}}
    \\
    -
    \bigg[
    \frac{
      \txonesq (1-\txtwo G_{\sigma n})
    }{
      (1-2\txtwo G_{\sigma n})^2
    }
    +
    \frac{\txtwosq }{(1-\txtwo G_{\sigma n})}
    \bigg]
    \,
    G_{\sigma n}
    \,.\label{eq:weiss-ttprime-hom}
  \end{multline}
  Returning to Eq.~(\ref{eq:weiss-ttprime-AB}) we note that for pure
  NN hopping ($\txtwo $ $=$ $0$) it reduces to the standard
  expression\cite{georges}
  \begin{align}
    \calG_{\gamma\sigma n}^{-1}
    &=
    \mathrm{i}\omega_n + \mu + h_{\gamma\sigma} - \txonesq G_{\bar{\gamma}n\sigma} 
    \,,\label{eq:weiss-NN}
  \end{align}
  while for pure NNN hopping ($\txone $ $=$ $0$)
  \begin{align}
    \calG_{\gamma\sigma n}^{-1}
    &= 
    \mathrm{i}\omega_n
    +
    \mu
    +
    h_{\gamma\sigma}
    -
    \frac{
      \txtwosq G_{\gamma\sigma n}
    }{
      1-\txtwo G_{\gamma\sigma n}
    }
    \,.\label{eq:weiss-NNN}
  \end{align}
  In the former case there is strong sublattice mixing since 
  hopping takes place only between different sublattices.
  In the latter case we find no mixing at all since hopping is
  allowed only between sites in the same sublattice.

  \section{Discussion}\label{sec:conclusion}
  
  On the Bethe lattice and the triangular Husimi cactus the number of
  possible paths joining two points is determined only by the
  topological distance between these points.  Using this property we
  derived the operator relation (\ref{eq:geometricseries}) and
  obtained the spectrum of arbitrary tight-binding Hamiltonians,
  without the need for complicated geometrical constructions of other
  methods.
  
  In Sec.~\ref{sec:dos} the density of states was calculated
  analytically for several classes of tight-binding Hamiltonians on
  the Bethe lattice.  As the NNN hopping amplitude increases it
  becomes asymmetric and develops a square-root singularity at a band
  edge. For the Hubbard model such a shape of the DOS is
  known\cite{wahle98,noack,vollhardt01} to support metallic ferromagnetism away from
  half-filling, and to suppress antiferromagnetism near half-filling.
  
  In Sec.~\ref{subsec:arbdos} we showed that for any given DOS one can
  determine the corresponding hopping parameters on a Bethe lattice.
  This result is useful in particular in the context of dynamical
  mean-field theory.  Namely, with this inverse construction it is now
  possible to mimic any kind of van-Hove singularity or other
  band structure even on the Bethe lattice.
  
  Furthermore, we derived the exact self-consistency equations of
  dynamical mean-field theory in Sec.~\ref{sec:dmft} \emph{for
    arbitrary hopping}.  In the particular case of $\tonetwo$ hopping
  these equations differ from those used in previous
  investigations,\cite{georges,rozenberg2,rozenberg,zitzler} which are
  therefore found to apply only to random hopping.  The exact DMFT setup
  derived in this paper [DOS (\ref{eq:rho-ttprime-infZ}) and Weiss
  field (\ref{eq:weiss-ttprime-AB})] can be expected to lead to new
  quantitative results for the phase diagram of the $\tonetwo$ Hubbard
  model on the Bethe lattice. This is clear from the exact relation
  (\ref{eq:weiss-ttprime-AB}) between the Weiss field and the
  interacting local Green function, since the Green function now
  appears also in the denominator, thus leading to resonances in the
  Weiss field. It will be most interesting to study the possible
  solutions obtained within this framework in detail.
  
  In conclusion, the method and results presented in this paper lay
  the foundation for systematic studies of correlated electronic
  systems on a Bethe lattice.  Possible future applications include
  geometrical frustration in itinerant systems or quantum magnets, and
  weak-localization effects in disordered systems.  Indeed, the
  relation (\ref{eq:geometricseries}) may also become quite useful for
  future investigations in statistical mechanics.

  \begin{acknowledgments}
    The authors thank N.\ Bl\"{u}mer, R.\ Bulla, G.\ Kotliar, and W.\ 
    Metzner for useful discussions. This work was supported in part by
    the SFB 484 of the Deutsche Forschungsgemeinschaft (DFG).
    Financial support of one of the authors (K.B.) through KBN-2 P03B
    08 224 is also acknowledged.
    
  \end{acknowledgments}

  \appendix

  \section{The triangular Husimi cactus}\label{sec:husimi}
  
\begin{figure}[t]
    \centerline{\includegraphics[clip,height=0.75\hsize,angle=0]{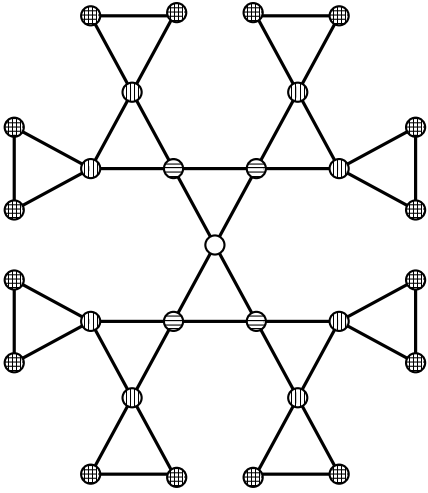}}
    \nopagebreak
    \caption{Part of the triangular Husimi cactus with $Z$ triangles 
      connected to each site; here $Z$ $=$ $2$.  Any two sites are
      connected by a unique shortest path of bonds.  The lattice is
      infinite and all sites are equivalent. The shading of sites has
      the same meaning as in Fig.~\ref{fig:bethe}.}
    \label{fig:husimi}
  \end{figure}

  The triangular Husimi cactus is a set of triangles connected in such
  a way that each vertex belongs to $Z$ triangles and each edge
  belongs to only one triangle (see Fig.~\ref{fig:husimi}).  Below we
  show how to express powers of the NN hopping Hamiltonian $H_1$ in
  terms of the hopping Hamiltonian $H_d$ between sites with
  topological distance $d$ [see Eq.~(\ref{eq:H-nu})].  For brevity we
  again set $K=Z-1$ and $p=Z/K$, and employ scaled operators
  $\tilde{H}_d$ $=$ $H_d/K^{d/2}$.

  One can check directly that the Husimi cactus is distance regular,
  i.e., the number of paths between two vertices $i$ and $j$ depends
  only on their topological distance. This shows that the absence of
  closed loops as on the Bethe lattice is \emph{not} a necessary
  condition for this property.  The number $a_n^{(d)}$ of paths of
  length $n$ between two points with distance $d_{ij}=d$ satisfies
  recursion relations very similar to those on the Bethe lattice: for
  $n,d$ $\geq 0$ we find $a_{n+1}^{(d+1)}$ $=$ $2K
  a_{n}^{(d+2)}+a_{n}^{(d+1)}+a_{n}^{(d)}$, $a_{n+1}^{(0)}$ $=$
  $2Za_{n}^{(1)}$, $a_0^{(d)}$ $=$ $\delta_{0d}$, and $a_n^{(n)}$ $=$
  $1$. Using Eq.~(\ref{eq:a-trafo}) we thus find for the first few
  powers of $H_1$
  \begin{subequations}%
  \begin{align}%
    (H_1)^2
    &=
    2Z\openone
    +
    H_1
    +
    H_2
    \,,\label{eq:husimi2}
    \\ 
    (H_1)^3
    &=
    2Z\openone
    +
    (4Z-1)H_1
    +
    2H_2
    +
    H_3
    \,,
    \\
    (H_1)^4
    &=
    (4Z-1)2Z\openone
    +
    5(2Z-1)H_1
    \nonumber\\&\phantom{=}\;
    +
    (6Z-1)H_2
    +
    3H_3
    +
    H_4
    \,.%
  \end{align}%
  \end{subequations}%
  Note that due to the presence of closed loops there are also paths
  of odd length joining vertices with even distance and vice versa.
  
  We now proceed similar to Sec.~\ref{sec:opident}. The recursion
  relations for the generating function
  $F_d(u)=\sum_{n=0}^{\infty}a_n^{(d)}u^n$ can again be solved by an
  ansatz\cite{riordan} $F_d(u)$ $=$ $f(u)\,[u\,g(u)]^d$. After some
  algebra we obtain the operator identity
  \begin{align}
    \frac{(1+2x)(1-x)}{1-x(H_1-1)+2Kx^2}
    &=
    \sum_{d=0}^{\infty} H_d x^d 
    \,,\label{eq:husimi-geometricseries}
  \end{align}
  relating nearest-neighbor and long-range hopping Hamiltonians on the
  Husimi cactus. (We omit explicit expressions for the coefficients
  $a_n^{(d)}$ or $A_n^{(d)}$.) Note that
  Eq.~(\ref{eq:husimi-geometricseries}) is somewhat more complicated
  than the corresponding relation for the Bethe lattice
  [Eq.~(\ref{eq:geometricseries})].
  
  The Green function (\ref{eq:resolvent}) can now be obtained directly
  from Eq.~(\ref{eq:husimi-geometricseries}), similar to
  Sec.~\ref{subsec:greenfunc}. We find
  \begin{align}
    G_{ij}(z)
    &=
    \bra{i}
    (z-\tone H_1)^{-1}
    \ket{j}
    =
    \frac{x^{d_{ij}}}{z-2xZ\tone }
    \,,\label{eq:husimi-green1}
  \end{align}
  where $x$ $=$ $[z/\tone -1+\sqrt{(z/\tone -1 )^2-8K}]/(4K)$ and $d_{ij}$
  again denotes the topological distance between sites $i$ and $j$.
  For $\tilde{H}_1$
  (i.e., $\tone =1/\sqrt{K}$) the DOS is given by
  \begin{align}
    \rho_1(\lambda)
    &=
    \delta_{K1}\frac{\delta(\lambda+2)}{3}
    +
    \frac{p}{2\pi}
    \frac{\sqrt{\smash[b]{8-(\lambda-1/\sqrt{K})^2}}}
    {(p+\lambda/\sqrt{K})(2p-\lambda/\sqrt{K})}
    \,,\label{eq:husimi-dos1}
  \end{align}
  implying  a continuous spectrum in the interval
  $1/\sqrt{K}-\sqrt{8}\leq \lambda \leq 1/\sqrt{K}+\sqrt{8}$ with an
  additional delta-peak in the case $K$ $=$ $1$.  Previous derivations
  of (\ref{eq:husimi-dos1}) can be found for $K$ $=$ $1$ in
  Ref.~\onlinecite{thorpe} and for $K$ $>$ $1$ in
  Ref.~\onlinecite{wahle97}. Note that Eq.~(\ref{eq:husimi-dos1})
  reduces to a semielliptical DOS for $Z$ $\to$
  $\infty$.
  
  As an application we now calculate the DOS for the Hamiltonian with
  $\tonetwo$ hopping, i.e.,
  \begin{align} 
    H_{\txone ,\txtwo}
    &=
    tH_1+\ttwo H_2
    =
    \txone\tilde{H}_1+\txtwo\tilde{H}_2
    \nonumber\\
    &=
    tH_1+\ttwo {[H_1^2-H_1-2Z\openone]}
    \,,
  \end{align}
  due to Eq. (\ref{eq:husimi2}). As before we use the scaling
  $\tone=\txone/\sqrt{K}$ and $\ttwo=\txtwo/K$ and find the
  eigenvalues $\epsilon$ of $H$ in terms of the eigenvalues $\lambda$
  of $\tilde{H}_1$, i.e.,
  \begin{align}
    \epsilon(\lambda)
    =
    \txtwo\lambda^2+(\txone -\txtwo/\sqrt{K})\lambda-2p\txtwo 
    \,,\label{eq:husimi-disp}
  \end{align}
  which has the roots
  $\lambda_{1,2}(\epsilon)
    =
    (\xi\pm\sqrt{\xi^2+\eta})/(2\txtwo)$, where
  \begin{align}
    \xi
    &=
    \frac{\txtwo}{\sqrt{K}}-\txone
    \,,~~
    \eta
    =
    4\txtwo(2p\txtwo+\epsilon)
    \,.
  \end{align}
  Finally the DOS is obtained by changing variables in
  Eq.~(\ref{eq:husimi-dos1}),
  \begin{align}
    \rho_{\txone ,\txtwo}(\epsilon)
    &=
    \delta_{K1}\frac{\delta(\epsilon+2(\txone -\txtwo ))}{3}
    +
    \sum_{i=1}^2
    \frac{\rho_1(\lambda_i(\epsilon))}{\sqrt{\xi^2+\eta}}
    \,.\label{eq:husimi-rho-ttprime}
  \end{align}
  The Husimi cactus is not a bipartite lattice, and the symmetry
  $\rho_{\txone ,\txtwo}(\epsilon)$ $=$ $\rho_{-\txone,\txtwo}(\epsilon)$ holds
  only for $K$ $=$ $\infty$; this can be seen from
  Eq.~(\ref{eq:husimi-disp}).  We omit plots of
  (\ref{eq:husimi-rho-ttprime}) since its behavior with varying $\txtwo $
  is rather similar to the DOS (\ref{eq:rho-ttprime}) for the Bethe
  lattice.

\end{document}